\documentclass[aps,prb,twocolumn,superscriptaddress,groupedaddress]{revtex4}
\usepackage[dvips,letterpaper,margin=0.5in,bottom=0.75in]{geometry} 
\usepackage[english]{babel}
\usepackage{epstopdf}
\usepackage{pdfpages}
\usepackage{float}
\usepackage{enumerate}
\usepackage{hyperref}

\usepackage{amsmath} 
\usepackage{dsfont}
\usepackage{cancel}
\usepackage{amsthm} 
\usepackage{amssymb}	
\usepackage{empheq}
\usepackage[absolute,overlay]{textpos}
\usepackage{xcolor}
\usepackage{color}

\allowdisplaybreaks

\DeclareMathOperator{\Tr}{Tr}

\begin{document}

\title{Cross-dimensional universality classes in static and periodically driven Kitaev models}


\date{\today}

\author{Paolo Molignini}
\affiliation{Clarendon Laboratory, University of Oxford, Parks Road, Oxford OX1 3PU, United Kingdom}
\affiliation{Cavendish Laboratory, University of Cambridge, 19 J J Thomson Avenue, Cambridge CB3 0HE, United Kingdom} 
\author{Albert Gasull Celades}
\affiliation{Max-Planck-Institut of Quantum Optics, Hans-Kopfermann-Str. 1, 85748 Garching, Germany}
\author{R. Chitra}
\affiliation{Institute for Theoretical Physics, ETH Z\"{u}rich, 8093 Zurich, Switzerland}
\author{Wei Chen}
\affiliation{Department of Physics, PUC-Rio, 22451-900 Rio de Janeiro, Brazil}

\begin{abstract}

The Kitaev model on the honeycomb lattice is a paradigmatic system known to host a wealth of nontrivial topological phases and Majorana edge modes.
In the static case, the Majorana edge modes are nondispersive. 
When the system is periodically driven in time, such edge modes can disperse and become chiral.
We obtain the full phase diagram of the driven model as a function of the coupling and the driving period.
We characterize the quantum criticality of the different topological phase transitions in both the static and driven model via the notions of Majorana-Wannier state correlation functions and momentum-dependent fidelity susceptibilities.
We show that the system hosts cross-dimensional universality classes: although the static Kitaev model is defined on a 2D honeycomb lattice, its criticality falls into the universality class of 1D linear Dirac models. 
For the periodically driven Kitaev model, besides the universality class of prototype 2D linear Dirac models, an additional 1D nodal loop type of criticality exists owing to emergent time-reversal and mirror symmetries, indicating the possibility of engineering multiple universality classes by periodic driving. 
The manipulation of time-reversal symmetry allows the periodic driving to control the chirality of the Majorana edge states.

\end{abstract}

\maketitle

\section{Introduction}

Majorana modes (MMs) are exotic non-abelian quasiparticles predicted to appear in topological superconductors~\cite{AliceaReview,Sato:2017, Aguado:2017}.
Besides representing solid-state realizations of the elusive Majorana fermion in particle physics~\cite{Majorana:1937}, their non-abelian exchange statistics could be exploited to robustly encode quantum information via braiding schemes~\cite{Kitaev:2003,Beenakker:2013,DasSarma:2015}.
In recent years, preliminary experimental evidence for their existence~\cite{Mourik:2012, Albrecht:2016, Deng:2016, Lutchyn:2018, ZhangNature:2018, Jaeck:2019, Manna:2020, Frolov:2020}  and notable progress in devising topological quantum computation protocols~\cite{Aasen:2016,Litinski:2018,Oreg:2020}, have fostered an intensive investigation of systems hosting MMs, such as the zero-dimensional (0D) edge state at the end of one-dimensional (1D) topological superconductors~\cite{Kitaev:2001}, and the 1D edge state at the boundaries of two-dimensional (2D) topological superconductors\cite{KitaevAnnals:2006,Nakosai13,Chen15_Majorana_multiferroics}.
If time-reversal symmetry (TRS) is present, as in the prototypical Kitaev model on a honeycomb lattice~\cite{KitaevAnnals:2006}, MMs are immobile (nondispersive).
For TRS breaking systems,  such as chiral $p$-wave superconductors~\cite{Read:2000,Chiu:2018, WangPRL:2018, He:2019}, MMs can propagate along the edges and acquire a chirality.
The mobility of chiral MMs may inherently facilitate quantum computing procedures~\cite{Lian:2018}.
However, TRS breaking typically requires the application of magnetic fields or complex pairing mechanisms with higher orbital angular momentum, which makes their solid-state realization challenging.

Recently, Floquet systems have emerged as a viable alternative to realize topological phases that are hard to fabricate in equilibrium, and also as a tool to explore MMs out of equilibrium~\cite{Jiang:2011,Liu:2013,Thakurathi:2013,Thakurathi:2014,Sacramento:2015, Bhattacharya:2016, Molignini:2017,Thakurathi:2017,Po:2017,Peng:2018,Molignini:2018,Cadez:2019,Molignini:2020-multifrequency}.
In particular, circular patterns of hoppings breaking TRS can be engineered to naturally pump topological modes along the boundaries of strip geometries~\cite{Rudner:2013, Po:2017, Mukherjee:2017, Mukherjee:2018, Yu:2020, Wintersperger:2020}.
Additionally, the periodicity in time results in Floquet-Brillouin zones which can potentially harbor two kinds of edge modes, leading to anomalous phases and cascades of nonequilibrium topological phase transitions (TPTs)~\cite{Rudner:2013, Klinovaja:2016, Mukherjee:2017, Mukherjee:2018, Molignini:2019, Wintersperger:2020}. Periodic driving can also induce topology beyond Dirac low-energy theories, \textit{e.g.} nodal loop semimetal phases~\cite{Li:2018, Molignini:2018}.
The richness of Floquet systems has spurred new classification schemes for nonequilibrium topology~\cite{Rudner:2015,Yao:2017,Harper:2017,Roy:2017,Harper:2020} and new protocols for  quantum computation~\cite{Bauer:2019}.

In this work, we aim to study TPTs in both the static and periodically driven 2D Kitaev models hosting MMs.
We classify these transitions from the perspective of universality using two measures:
i) a Majorana version of a recently proposed stroboscopic Wannier state correlation function~\cite{Chen:2017,Molignini:2018,Chen-Sigrist-book:2019,Chen-Schnyder:2019}, 
which encodes the notion of a correlation length that diverges at the TPT and ii) a momentum-dependent fidelity susceptibility that measures the distance between Bloch states in the momentum space~\cite{Panahiyan20}. 
Our classification reveals the existence of \emph{cross-dimensional} universality classes.
In the static model, despite being defined on a 2D honeycomb lattice, the transitions in fact belong to the universality class of 1D Dirac models~\cite{Chen:2017,Chen-Sigrist-book:2019,Chen-Schnyder:2019}.
In the presence of periodic driving, we show that TRS can be manipulated and consequently help to switch between immobile and chiral MMs.
Owing to the ability to manipulate the symmetry of the system, the periodically driven case displays an even richer criticality with two universality classes of different dimensionality: one representing prototype 2D linear Dirac models, and another corresponding to a 1D nodal loop semimetal type due to emergent time-reversal and mirror symmetries.

This paper is structured as follows.
In Sec.~\ref{sec:model}, we introduce the Majorana-Wannier state correlation function and fidelity susceptibility to delineate the quantum criticality in the static Kitaev model. 
In Sec.~\ref{sec:driven-topo},  we discuss the phase diagram of the  periodically driven Kitaev model and the emergence of chiral MMs in the quasienergy spectrum. 
We quantify the criticality  of the transitions seen using the notions of stroboscopic Wannier state correlation function and the fidelity susceptibility.
In Sec.~\ref{sec:NL}, we focus on a nodal loop semimetal phase arising from frozen dynamics, address the corresponding emergent symmetries, and determine its universality class.
Finally, Sec.~\ref{sec:outlook} concludes our discussion with an outlook for future research.

\newpage
\onecolumngrid

\begin{figure}
\centering
\includegraphics[width=\textwidth]{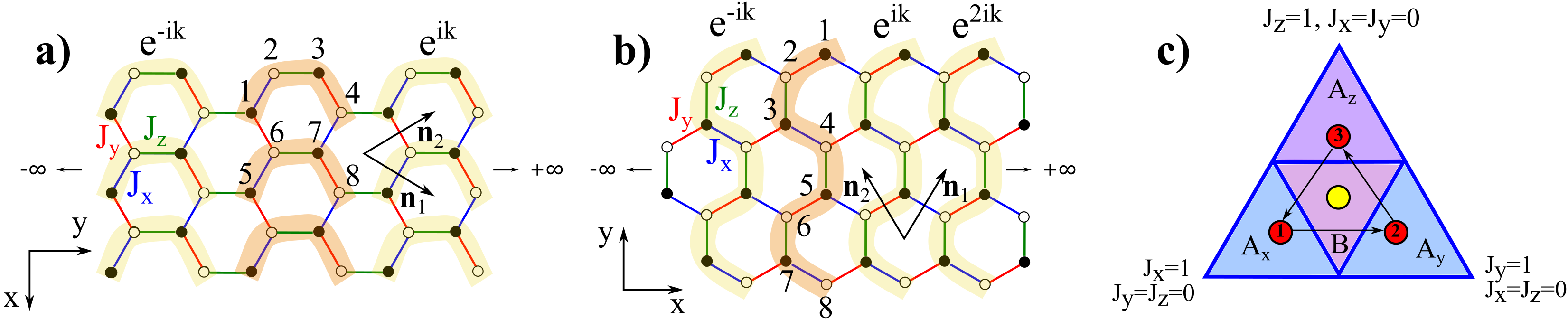}
\caption{Sketch of the Kitaev model for a semi-infinite strip with a) armchair and b) zigzag edges. The red shaded regions and the numbering denote the superunit cell, while the yellow shaded regions mark the corresponding sites obtained by translations by multiples of the momentum $k$. c) Static phase diagram in the $(J_x, J_y, J_z)$ space and illustration of the driving protocol (red circles). At the special point $J=1/3$ (yellow circle), all the driving steps are equal.}
\label{fig:sketch}
\end{figure}

\twocolumngrid

\section{Static Kitaev model}
\label{sec:model}

We consider the 2D Kitaev model, consisting of spin-$1/2$'s localized at the vertices of a honeycomb lattice spanned by basis vectors $\mathbf{n}_1 = \left( \frac{1}{2}, \frac{\sqrt{3}}{2} \right)$ and  $\mathbf{n}_2 = \left( -\frac{1}{2}, \frac{\sqrt{3}}{2} \right)$.
The system is described by the following Hamiltonian~\cite{KitaevAnnals:2006, Thakurathi:2014}
\begin{equation}
\mathcal{H} = J_x \sum_{\text{x links}} \sigma_j^x \sigma_k^x  + J_y \sum_{\text{y links}} \sigma_j^y \sigma_k^y + J_z \sum_{\text{z links}} \sigma_j^z \sigma_k^z,
\end{equation}
where $\sigma_k^i$ is the Pauli matrix representing the $i$-th spin component at site $k$, while $J_i \ge 0$ is the coupling strength of the $i$-type bond.
A physical implementation of this model has been proposed in cold polar molecules~\cite{Micheli:2006,Gorshkov:2013}.

To understand the physics of this model,  note that for every hexagonal plaquette in the lattice, there exists a conserved quantity $W_p = \sigma_1^x  \sigma_2^y  \sigma_3^z  \sigma_4^x  \sigma_5^y  \sigma_6^z$ which commutes with the Hamiltonian.   
The eigenstates of the system therefore, split into two sectors denoted by the eigenvalues $w_p= \pm 1$ of $W_p$. 
Throughout our analysis, we will exclusively consider the vortex-free sector $w_p=+1$, wherein the ground state  lies~\cite{KitaevAnnals:2006}.
The vortex-free sector can be solved exactly by mapping the spins onto Majorana operators~\cite{Baskaran:2007,Lee:2007,Chen:2008,Nussinov:2008,Thakurathi:2014} to yield the Hamiltonian
\begin{equation}
\mathcal{H} = i \sum_{\mathbf{n}} \left( J_x b_{\mathbf{n}} a_{\mathbf{n} - \mathbf{n}_1} +  J_y b_{\mathbf{n}} a_{\mathbf{n}  + \mathbf{n}_2} + J_z b_{\mathbf{n}} a_{\mathbf{n}} \right),
\label{Majorana-Ham}
\end{equation}
where the Majorana operators satisfy the anticommutation relations $\{ a_{\mathbf{n}}, a_{\mathbf{m}} \} = 2 \delta_{n_x m_x} \delta_{n_y m_y}$,  $\{ b_{\mathbf{n}}, b_{\mathbf{m}} \} = 2 \delta_{n_x m_x} \delta_{n_y m_y}$, and $\{ a_{\mathbf{n}}, b_{\mathbf{m}} \} =0$.
To probe the edge states appearing in this model and its driven counterpart, we will consider two kinds of edge geometries for semi-infinite honeycomb lattice strips, namely zigzag and armchair edges~\cite{Nakada:1996, Kohmoto:2007}. This amounts to rewriting \eqref{Majorana-Ham} as~\cite{Thakurathi:2014}
\begin{equation}
\mathcal{H} = i \sum_{j,l} {\rm w}_j M_{jl}(k) {\rm w}_l, 
\label{matrix-Ham}
\end{equation}
where the vector $\bf{w}$ collects the Majorana operators at different sites following the numbering defined by the superunit cell, and $M(k)$ is a matrix that describes the lattice connectivity in the finite direction for every momentum $k$ in the infinite direction. The two  geometries and the corresponding super-unit cells used in our construction of the matrix Hamiltonian $M$ are depicted in Fig.~\ref{fig:sketch}.

We will also benchmark our results by considering a system with periodic boundary conditions (or infinite) in both directions.
In this case, the bulk Hamiltonian can be written in Fourier space as~\cite{Thakurathi:2014}
\begin{equation}
\mathcal{H}_{\text{2D Kitaev}} = \sum_{\mathbf{k} \in \frac{1}{2} \text{BZ}} \left( \begin{array}{cc} a_{-\mathbf{k}} & b_{-\mathbf{k}} \end{array} \right) \mathcal{H}(\mathbf{k}) 
\left( \begin{array}{c} a_{\mathbf{k}} \\ b_{\mathbf{k}} \end{array} \right) 
\label{Majorana-rep-Kitaev-honeycomb}
\end{equation}
by defining Fourier transforms
\begin{align}
a_{\mathbf{n}} = \sqrt{\frac{4}{N}} \sum_{\mathbf{k} \in \frac{1}{2} \mathrm{BZ}} \left( a_{\mathbf{k}} e^{i \mathbf{k} \cdot \mathbf{n}} + a_{-\mathbf{k}} e^{-i\mathbf{k} \cdot \mathbf{n}} \right) \\
b_{\mathbf{n}} = \sqrt{\frac{4}{N}} \sum_{\mathbf{k} \in \frac{1}{2} \mathrm{BZ}} \left( b_{\mathbf{k}} e^{i \mathbf{k} \cdot \mathbf{n}} + b_{-\mathbf{k}} e^{-i\mathbf{k} \cdot \mathbf{n}} \right),
\end{align}
with $\{ a_{\mathbf{k}}, a_{\mathbf{k}'} \}= \{ b_{\mathbf{k}}, b_{\mathbf{k}'} \} = \delta(\mathbf{k} - \mathbf{k}')$ and 
\begin{align}
\mathcal{H}(\mathbf{k}) &=  2 \left[ J_x \sin \left( \mathbf{k} \cdot \mathbf{n}_1 \right) + J_y \sin \left( \mathbf{k} \cdot \mathbf{n}_2 \right) \right]  \sigma^x + \nonumber \\
& \quad + 2 \left[ J_x \cos \left( \mathbf{k} \cdot \mathbf{n}_1 \right) + J_y \cos \left( \mathbf{k} \cdot \mathbf{n}_2 \right) + J_z \right]  \sigma^y
\nonumber \\
&=d_{1}\sigma^{x}+d_{2}\sigma^{y}.
\label{bulk-Ham}
\end{align}
The corresponding eigenstates and eigenenergies are
\begin{align}
|u_{\pm}({\bf k})\rangle &=\frac{1}{\sqrt{2}d}\left(\begin{array}{c}
\pm d \\
d_{1}+id_{2}
\end{array}\right),
\nonumber \\
E_{\pm}(\mathbf{k}) &= \pm 2 \big( \left[  J_x \sin \left( \mathbf{k} \cdot \mathbf{n}_1 \right) + J_y \sin \left( \mathbf{k} \cdot \mathbf{n}_2 \right) \right]^2  \nonumber \\
& \qquad \quad \left[ J_x \cos \left( \mathbf{k} \cdot \mathbf{n}_1 \right) + J_y \cos \left( \mathbf{k} \cdot \mathbf{n}_2 \right) + J_z \right]^2  \big)^{1/2},
\label{Kitaev-honeycomb-disp}
\end{align}
where $d=\sqrt{d_{1}^{2}+d_{2}^{2}}$. Note that the sum in Eq.~\eqref{Majorana-rep-Kitaev-honeycomb} runs over only half of the Brillouin zone. 
A convenient choice of the Brillouin zone consists of the rectangle given by $k_x \in [-2\pi, 2\pi]$ and $k_y \in [ \frac{2\pi}{\sqrt{3}}, -\frac{2\pi}{\sqrt{3}}]$.

Following Ref.~\cite{Thakurathi:2014}, the phase diagram can be deduced from Eq.~\eqref{Kitaev-honeycomb-disp} by determining the parameters where  gap closures occur. 
These points fulfil the condition
\begin{equation}
J_x \le J_y + J_z, \quad J_y \le J_x + J_z, \quad J_z \le J_y + J_x,
\end{equation}
Setting $J_x + J_y + J_z = 1$, the phase diagram can be depicted as an equilateral triangle comprizing   four distinct topological sectors  $A_x$, $A_y$, $A_z$, and $B$,  see Fig.~\ref{fig:sketch}c).
The $A$ phases are all gapped ($E(\mathbf{k})\neq 0, \forall \mathbf{k}$), while the $B$ phase is gapless.
Static zero-energy MMs exist in the regions $A_z$ and $B$ for zigzag edges, and in the regions $A_x$, $A_y$ and $B$ for armchair edges~\cite{Thakurathi:2014}.

\subsection{Correlation function and fidelity susceptibility for static Kitaev model}
\label{sec:static_Majorana_Wannier_fidelity}

The time-reversal symmetric static Kitaev model belongs to the 2D class DIII, whose topological invariant is determined by the Pfaffian of the matrix elements of the time-reversal operator $m_{\alpha\beta}({\bf k})=\langle u_{\alpha}({\bf k})|{\cal T}|u_{\beta}({\bf k})\rangle$, where $|u_{\alpha}({\bf k})\rangle$ is the Bloch eigenstate in Eq.~(\ref{Kitaev-honeycomb-disp}). To be specific, one integrates the derivative of the logarithm of the Pfaffian over the boundary of the half-Brillouin zone (hBZ) to construct the topological invariant 
\begin{eqnarray}
\nu=\frac{1}{2\pi i}\int_{\partial\frac{1}{2}BZ}d\log\left[{\rm Pf}(m)\right]=\frac{1}{2\pi}\int_{\partial\frac{1}{2}BZ}d{\bf k}\cdot{\boldsymbol\nabla}\phi,
\label{topo_inv_static_Kitaev}
\end{eqnarray}
which is equivalently the winding number of the phase $\phi=\arctan(d_{2}/d_{1})$ of the off-diagonal element of the Hamiltonian in Eq.~(\ref{bulk-Ham}) along the hBZ boundary. 

\begin{figure}
\centering
\includegraphics[width=\columnwidth]{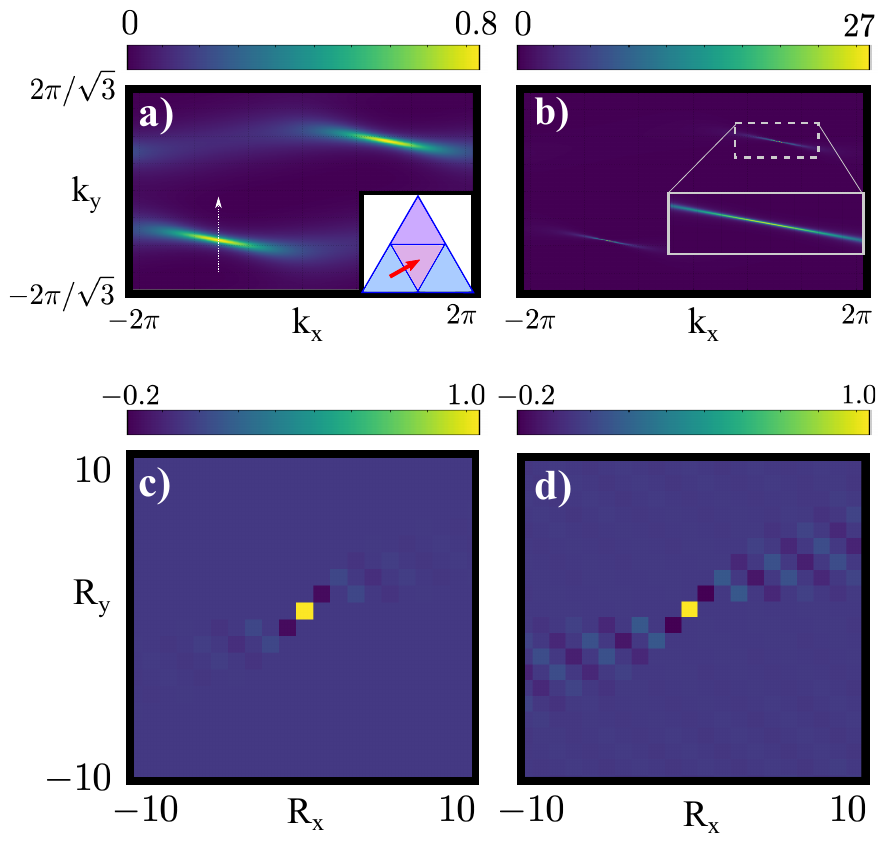}
\caption{a), b) An example of the critical behavior of the curvature function $F({\bf k},{\bf J})$ approaching the TPT which shows a Lorentzian shape along certain direction (here for the $A_x$-$B$ TPT shown in the inset, the other two TPTs show analogous behaviors but rotated by 60 degrees in $\mathbf{k}$-space). 
The Lorentzian shape gradually narrows and then flips sign as the system crosses the TPT. 
The dotted vertical arrow indicates a possible scaling direction $\mathbf{k}_s$ used in the extraction of the critical exponents.
c)-d) The Fourier transform of the curvature function (here calculated as the fast Fourier transform of the curvature function), which represents a Majorana-Wannier state correlation function, decays in real space with a correlation length $\xi$ that diverges at the critical point.
The Majorana-Wannier state correlation function is also strongly aligned along one direction, confirming 1D behavior.} 
\label{fig:static_Kitaev_FkM_corr}
\end{figure}

Our aim is to introduce a correlation function to describe the quantum criticality near the TPTs caused by the tuning parameters ${\bf J}=(J_{x},J_{y},J_{z})$. For this purpose, we consider the gradient of the phase in Eq.~(\ref{topo_inv_static_Kitaev}) along a specific scaling direction ${\hat{\bf k}}_{s}$ to define a curvature function~\cite{vonGersdorff:2021} 
\begin{eqnarray}
&F({\bf k},{\bf J})={\hat{\bf k}}_{s}\cdot{\boldsymbol\nabla}\phi=\cos\beta\,\partial_{x}\phi+\sin\beta\,\partial_{y}\phi\equiv\partial_{s}\phi
\nonumber \\
&=\frac{d_{1}\partial_{s}d_{2}-d_{2}\partial_{s}d_{1}}{d^{2}}=-2\langle u_{-}|i\partial_{s}|u_{-}\rangle,
\label{FkM_Berry_con_static_Kitaev}
\end{eqnarray}
after using the eigenstate in Eq.~(\ref{Kitaev-honeycomb-disp}), which is essentially the integrand in Eq.~(\ref{topo_inv_static_Kitaev}).
There is a degree of freedom in Eq.~(\ref{topo_inv_static_Kitaev}) corresponding to the choice of the path of the contour integral, which is reflected in the freedom of choosing the scaling direction $\mathbf{k}_s$ and the corresponding curvature function \eqref{FkM_Berry_con_static_Kitaev}.
For simplicity, here we choose the contour such that it crosses all the HSPs along the same direction $\mathbf{k}_s = \hat{\mathbf{k}}_y$, generating the same curvature function at all HSPs.
We remark that even though different choices for $\mathbf{k}_s$ are possible, all of them yield the same critical lines and exponents.

Equation (\ref{FkM_Berry_con_static_Kitaev}) indicates the equivalence between the curvature function in this problem and the Berry connection of the filled band eigenstate. 
Moreover, we find that close to TPTs, the curvature function exhibits diverges at the HSPs ${\bf k}_{0}= \pm (\pi, \frac{\pi}{\sqrt{3}})$, $\pm (-\pi, \frac{\pi}{\sqrt{3}})$, $\pm (0, \frac{2\pi}{\sqrt{3}})$, corresponding to the $A_x$-$B$,  $A_y$-$B$, and $A_z$-$B$ TPTs, respectively.
These HSPs can be obtained analytically by calculating $F(\mathbf{k}, \mathbf{J})$ at the corresponding values of the hopping parameters, and setting its denominator to zero.
Expanding along the scaling direction $\delta{\bf k}_{s}={\hat{\bf k}}_{s}\delta k_{s}$, the  divergent satisfy an Ornstein-Zernike form
\begin{eqnarray}
F({\bf k}_{0}+\delta {\bf k}_{s},{\bf J})=\frac{F({\bf k}_{0},{\bf J})}{1+\xi^{2}\delta k_{s}^{2}}.
\label{static_Ornstein_Zernike}
\end{eqnarray}
As the system approaches the TPT ${\bf J}\rightarrow{\bf J}_{c}$, the Lorentzian shape gradually narrows and diverges, and flips sign as the system crosses the critical point.
Mathematically,
\begin{eqnarray}
&&\lim_{{\bf J}\rightarrow{\bf J}_{c}^{+}}F({\bf k}_{0},{\bf J})=-\lim_{{\bf J}\rightarrow{\bf J}_{c}^{-}}F({\bf k}_{0},{\bf J})=\pm\infty,
\nonumber \\
&&\lim_{{\bf J}\rightarrow{\bf J}_{c}}\xi=\infty,
\label{Fk0_xi_divergence}
\end{eqnarray}
which then defines the  critical exponents  $\gamma$ and $\nu$ as
\begin{eqnarray}
F({\bf k}_{0},{\bf J})\propto|{\bf J}-{\bf J}_{c}|^{-\gamma},\;\;\;\xi\propto|{\bf J}-{\bf J}_{c}|^{-\nu}.
\label{Fk0_xi_exponent}
\end{eqnarray}
An example of this critical behavior is shown in Fig.~\ref{fig:static_Kitaev_FkM_corr}. 
This critical behavior prompts us to investigate the Fourier transform of the curvature function.
Given the Wannier state localized at home cell ${\bf R}$ constructed from the filled band Bloch state in Majorana basis $\left| u_-(\mathbf{k})\right>$~\cite{Molignini:2018}, \textit{i.e.}
\begin{eqnarray}
|{\bf R}\rangle=\frac{1}{N}\sum_{\bf k}e^{i{\bf k}\cdot({\hat{\bf r}}-{\bf R})}|u_{-}({\bf k})\rangle,
\label{Wannier_static}
\end{eqnarray}
the Fourier transform of the curvature function represents the  following Wannier expectation value of the position operator ${\hat{\bf k}}_{s}\cdot{\hat{\bf r}}$ 
\begin{eqnarray}
&\tilde{F}_{st}({\bf R},{\bf J})=\int\frac{d^{2}{\bf k}}{(2\pi)^{2}}F({\bf k},{\bf J})e^{i{\bf k\cdot R}}=-2\langle {\bf R}|{\hat{\bf k}}_{s}\cdot{\hat{\bf r}}|{\bf 0}\rangle
\nonumber \\
&=-2\int d^{2}{\bf r}\,{\hat{\bf k}}_{s}\cdot{\bf r}\,W^{\ast}({\bf r-R})W({\bf r}).
\end{eqnarray}
Because the Wannier state is written in the Majorana basis, we call this overlap a Majorana-Wannier state correlation function. 
The Ornstein-Zernike form in Eq.~(\ref{static_Ornstein_Zernike}) further dictates that this correlation function decays with the correlation length $\xi$, as shown for several parameters in Fig.~\ref{fig:static_Kitaev_FkM_corr}, justifying the interpretation of $\nu$  as a correlation exponent.

The critical exponent  $\gamma$ is, on the other hand,  linked to the fidelity susceptibility~\cite{You07,Zanardi07,Gu10}.
 This can be seen using the quantum metric formalism~\cite{Provost80,Berry89}. 
As opposed to the approach used in Refs.~\cite{Yang08_Kitaev,Gu09_Kitaev,Wang10_Kitaev,Mukherjee12_Kitaev}, here we follow the approach of Ref.~\cite{Panahiyan20}, which demonstrated that a fidelity susceptibility that describes the evolution of Bloch eigenstate in momentum space can better interpret the scaling law between the exponents $\gamma$ and $\nu$. 
To do this, we consider the  change in the filled band Bloch eigenstate $|u_{-}({\bf k})\rangle$  under a  momentum shift ${\bf k}$ to ${\bf k}+\delta{\bf k}_{s}$ along the scaling direction. 
The rotation of $|u_{-}({\bf k})\rangle$ in the Hilbert space along this trajectory can be described by the overlap $|\langle u_{-}({\bf k})|u_{-}({\bf k}+\delta{\bf k}_{s})\rangle|=1-g_{ss}\delta k_{s}^{2}/2$,
which defines the quantum metric 
\begin{eqnarray}
g_{ss}&=&\langle\partial_{s}u_{-}|\partial_{s}u_{-}\rangle-\langle\partial_{s}u_{-}|u_{-}\rangle\langle u_{-}|\partial_{s}u_{-}\rangle.
\end{eqnarray}
For the Hamiltonian in Eq.~(\ref{bulk-Ham}) of the form ${\cal H}({\bf k})=d_{1}\sigma^{x}+d_{2}\sigma^{y}$, it can be verified that\cite{Panahiyan20}
\begin{eqnarray}
g_{ss}=\left[\langle u_{-}|i\partial_{s}|u_{-}\rangle\right]^{2}=\frac{1}{4}\partial_{s}\hat{\bf d}\cdot\partial_{s}\hat{\bf d}=\frac{1}{4}F({\bf k},{\bf J})^{2}\equiv\chi_{F},
\label{static_Kitaev_g_chi}
\end{eqnarray}
with ${\hat{\bf e}}_{s}=\partial_{s}\hat{\bf d}/2$ playing the role of a vierbein. 
We have introduced the fidelity susceptibility $\chi_{F}$ by regarding $|u_{-}({\bf k})\rangle$ as a two-component spinor and the momentum along the scaling direction $\delta{\bf k}_{s}$ as the tuning parameter. 
From Eq.~(\ref{static_Kitaev_g_chi}), we note that the fidelity susceptibility is governed by the same exponent as the curvature function $F({\bf k}_{0},{\bf J})$ at the HSP, justifying the nomenclature of a susceptibility exponent $\gamma$ in Eq.~(\ref{Fk0_xi_exponent}). 
Physically, the divergence of $\chi_{F}$ means the eigenstate at the HSP $|u_{-}({\bf k}_{0})\rangle$ and  slightly away along the scaling direction $|u_{-}({\bf k}_{0}+\delta{\bf k}_{s})\rangle$ become more and more orthogonal. 

A straight forward expansion centered around the HSPs yields the following critical behavior for the curvature and the correlation length:
\begin{equation}
|F(\mathbf{k}_0, \mathbf{J})| = \begin{cases}
\frac{-(J_x+J_y) \cos \beta + \sqrt{3} (-J_x + J_y) \sin \beta}{2(J_x - J_y - J_z)}, \: &\scriptstyle{\text{$A_x$-$B$ TPT}} \\
\frac{-(J_x+J_y) \cos \beta + \sqrt{3} (-J_x + J_y) \sin \beta}{2(J_x - J_y + J_z)}, \: &\scriptstyle{\text{$A_y$-$B$ TPT}} \\
\frac{-(J_x+J_y) \cos \beta - \sqrt{3} (J_x + J_y) \sin \beta}{2(J_x + J_y - J_z)}, \: &\scriptstyle{\text{$A_z$-$B$ TPT}}
\end{cases}
\label{eq:suscept}
\end{equation}
\begin{equation}
\xi^2 = \begin{cases}
\frac{f_1(\mathbf{J}, \beta)}{16(J_x-J_y-J_z)^2 ((J_x+J_y)\cos \beta + \sqrt{3} (J_x - J_y) \sin \beta)},  \: &\scriptstyle{\text{$A_x$-$B$ TPT}} \\
\frac{f_2(\mathbf{J}, \beta)}{16(J_x-J_y+J_z)^2 ((J_x+J_y)\cos \beta + \sqrt{3} (J_x - J_y) \sin \beta)},  \: &\scriptstyle{\text{$A_y$-$B$ TPT}} \\
\frac{f_2(\mathbf{J}, \beta)}{16(J_x+J_y-J_z)^2 ((J_x-J_y)\cos \beta + \sqrt{3} (J_x + J_y) \sin \beta)},  \: &\scriptstyle{\text{$A_z$-$B$ TPT}}
\end{cases}
\label{eq:xi}
\end{equation}
where $f_i(\mathbf{J}, \beta)$ are regular, non-diverging functions and $\beta$ the corresponding angle of $k_s$ on the cartesian plane. 
Note that the divergence of the $\beta$-dependent part of the denominators of Eq. \eqref{eq:xi} cancels out with the zeros of Eq. \eqref{eq:suscept}, and therefore does not correspond to critical behavior.
The divergence is exclusively governed by $J$-dependent factors in the denominators.
Extracting the critical exponents from Eqs. \eqref{eq:suscept} and \eqref{eq:xi}, we obtain
\begin{eqnarray}
\gamma=\nu=1,
\end{eqnarray}
for all the TPTs in the static Kitaev model.  The exponents satisfy the scaling law $\gamma=\nu$  reminiscent of TPTs in 1D systems\cite{Chen:2017,Chen-Sigrist-book:2019,Chen-Schnyder:2019}  implying  that although the Kitaev model is defined on a 2D honeycomb lattice, it belong to the universality class of 1D linear Dirac models. 
This  is reflected  in Fig.~\ref{fig:static_Kitaev_FkM_corr}, where the curvature function near the HSP has a very elongated  quasi-1D shape. 
Moreover, this is compatible with the observation that the 2D Kitaev model on semi-infinite geometries in the vortex free sector can be mapped to a family of 1D Kitaev-like chains~\cite{Thakurathi:2014}.
In what follows, we will show that the system shows a far richer range of behaviors when it is periodically driven.

\section{Periodically driven Kitaev model}
\label{sec:driven-topo}

We will now consider a periodic modulation of the coupling parameters in a three-step fashion:
\begin{align}
J_i(t) = \begin{cases} 
J, \qquad & \frac{i-1}{3}T \le t \le \frac{i}{3} T \\
\frac{1-J}{2} \qquad & \text{else}
\end{cases}
\label{eq:driving-scheme}
\end{align}
with $J \in [0,1]$ and period $T$, and where for convenience we have relabelled $J_1 = J_x$, $J_2 = J_y$, and $J_3 = J_z$.
The driving scheme is applied after having fixed the system to be in the vortex-free sector and is not expected to lead to a mixing with the vortex-full sector because the two sectors are disjoint.
In parameter space, this modulation traces a typical path connecting the red dots as depicted in  Fig.~\ref{fig:sketch}c).
Similar piecewise modulations in different lattices have been explored in theoretical~\cite{Rudner:2013,Molignini:2019} and experimental works~\cite{Mukherjee:2017,Mukherjee:2018, Wintersperger:2020}, especially in the context of Floquet anomalous topological edge modes.
In particular, the scheme of \eqref{eq:driving-scheme} in the strong driving limit $J=1$ and fixed $T$ was shown to lead to radical chiral Floquet phases characterized by excitations with fractional statistics~\cite{Po:2017}.
Another work~\cite{Fulga:2019} analyzed a different four-step modulation and the stability of the chiral MMs with respect to disorder.

Here, we explore the full stroboscopic phase diagram as a function of \emph{both} $T$ and the full range of $J \in [0,1]$.
The phase diagram reveals many different phases, with a variable number of zero- and $\pi$-Majorana edge modes, including anomalous phases.
Through driving, the immobile Majorana edge modes present in the static model become both {\it propagating and chiral}. 
We  show that  velocity of the chiral modes can be  tuned by   varying $J$ and $T$.
Additionally,  tuning across $J=\frac{1}{3}$, we obtain a cascade of nodal loop gap closures that emerge because of additional symmetries effectively induced by the driving procedure.

To analyze the driven system, we first construct the Floquet operator~\cite{Thakurathi:2014} $U_F \equiv U(t=T)$ from the time evolution operator~\footnote{For simplicity, we will set $\hbar=1$ and the starting time of every time integration at zero throughout this paper.}
\begin{equation}
U(t) \equiv \mathbb{T} \left[ \exp \left( 4 \int_0^t \mathrm{d}t \: M(t) \right) \right],
\end{equation}
where $\mathbb{T}$ indicates time ordering and $M(t)$ is the Hamiltonian matrix describing the given geometry in the Majorana basis in Eq. \eqref{matrix-Ham}, but with time-dependent parameters $J_i(t)$ (the $k$-dependence has been omitted to highlight the fact that this procedure can be used also on geometries that are finite in both directions).
For our piecewise constant three-step driving protocol, the Floquet operator can be evaluated analytically as a product of matrix exponentials.
The effective Floquet Hamiltonian defined as ~\cite{}
\begin{equation}
\mathcal{H}_{\text{eff}} \equiv \frac{i}{T} \log U_F.
\label{eff-Floquet-Ham}
\end{equation}
 contains the full information about the system at stroboscopic times $t=N T$.
Diagonalization of $\mathcal{H}_{\text{eff}} T$ yields the  quasienergy spectrum $\epsilon_{\alpha}$ of the Floquet-state solutions $\Psi_{\alpha}(t) = \exp(-i\epsilon_{\alpha} t) \Phi_{\alpha}(t)$, where $\Phi_{\alpha}(t) = \Phi_{\alpha}( t+ T)$~\citep{Dittrich:1998}. 
Because of the $T$-periodicity of the Floquet modes $\Phi_{\alpha}(t)$, the quasienergies are restricted to lie in the interval $[-\pi, \pi)$.

To map out the driven topology, we consider the Floquet effective bulk Hamiltonian $\mathcal{H}_{\text{eff}} (\mathbf{k}) \equiv \frac{i}{T} \log U_F(\mathbf{k})$, with $U_F(\mathbf{k}) \equiv U(\mathbf{k} ; t=T) = \mathbb{T} \left[ \exp \left( -i \int_0^T \mathrm{d} t \: \mathcal{H}(\mathbf{k}; t) \right) \right]$, the equivalent Floquet  evolution operator defined for the bulk Hamiltonian \eqref{bulk-Ham}.
For the multistep drive, the resulting Floquet operator takes the form:
\begin{equation}
U_F(\mathbf{k}) = \left( \begin{array}{cc}
A(\mathbf{k}) & B(\mathbf{k}) \\
-B^*(\mathbf{k}) & A^*(\mathbf{k}).
\end{array}
\right)
\end{equation}
The analytic expressions for $A$ and $B$ being cumbersome are omitted here and instead presented in appendix \ref{app:anal-UF}.
Correspondingly, the effective Hamiltonian is given by 
\begin{align}
{\tiny
\mathcal{H}_{\text{eff}}(\mathbf{k}) } &= {\tiny \Lambda \left[ \Im[B(\mathbf{k})] \sigma^x + \Re[B(\mathbf{k})] \sigma^y + \Im[A(\mathbf{k})] \sigma^z \right]} \nonumber \\
&\equiv \mathbf{d}(\mathbf{k}) \cdot \boldsymbol{\sigma},
\label{Heff_strobo}
\end{align}
where $\Lambda(\mathbf{k}) =  \frac{\log \lambda^-(\mathbf{k}) - \log \lambda^+(\mathbf{k}) }{\lambda^+(\mathbf{k})  - \lambda^-(\mathbf{k}) }$ and $\lambda^{\pm}(\mathbf{k}) $ are the eigenvalues of $U_F(\mathbf{k})$.
The quasienergy dispersion is calculated as~\cite{Molignini:2019, Molignini:2020-multifrequency}
\begin{equation}
\theta(\mathbf{k}) = \arccos \left( \Tr[ U_F(\mathbf{k})/2] \right).
\label{eq:quasienergy-disp}
\end{equation}
Since the only symmetry fulfilled by the effective bulk Hamiltonian is charge conjugation, the stroboscopic topological invariant is the Chern number~\cite{ChiuReview:2016,vonGersdorff:2021}
\begin{equation}
C = -\frac{1}{2\pi} \int_{\text{BZ}} \: \mathrm{d}^2 k \: F(\mathbf{k}),
\label{top-inv-strobo}
\end{equation}
with the stroboscopic Berry curvature $F(\mathbf{k}) =  \hat{\mathbf{d}} \cdot \left[ \frac{\partial \hat{\mathbf{d}}}{\partial  k_x} \times \frac{\partial \hat{\mathbf{d}}}{\partial  k_y} \right]$ and $\hat{\mathbf{d}}= \frac{\mathbf{d}}{|\mathbf{d}|}$.

\subsection{Phase diagram}
\label{subsec:PD}

\begin{figure}
\centering
\includegraphics[width=0.9\columnwidth]{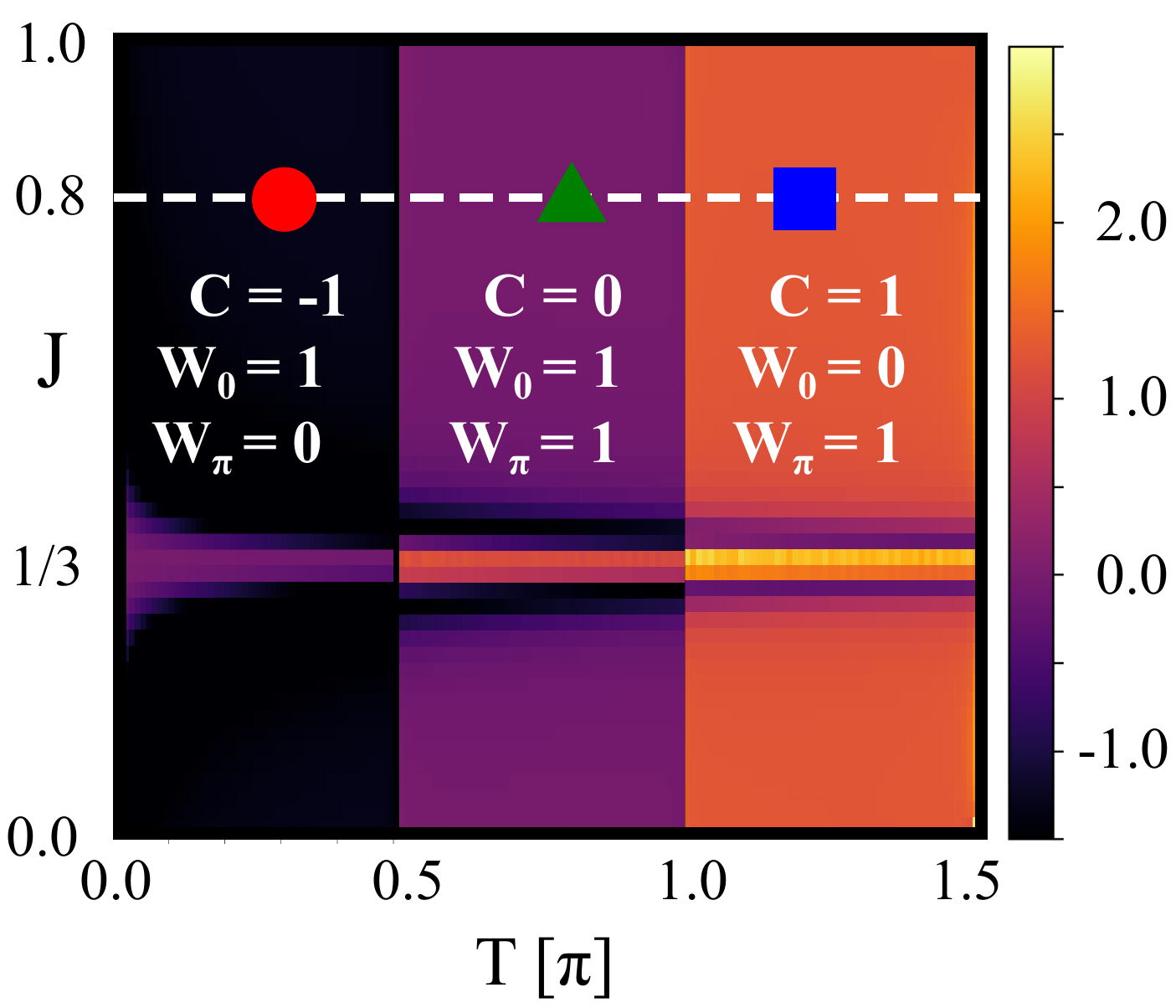}
\caption{Topological phase diagram in the ${\bf M}=(J,T)$ parameter space as computed from the topological invariant $\mathcal{C}$.
To verify the nature of each phase (\textit{e.g.} anomalous or not), the time-integrated invariants $W_0$ and $W_{\pi}$ were also calculated for representative points in each phase.
The colored symbols indicate the locations of exemplary quasienergy spectra and eigenmodes plotted in Figs.~\ref{fig:chiral-quasienergies} and \ref{fig:chiral-eigenfunctions}: $J=0.8$ and $T=0.3\pi$ (red circle), $T=0.7\pi$ (green triangle), $T=1.2\pi$  (blue square).
The stroboscopic invariant at each point was calculated for a grid of $200 \times 115$ points in $\mathbf{k}$-space.
} 
\label{fig:phase-diagram}
\end{figure}

 Our results for the topological phase diagram of the driven Kitaev model as a function of $J$ and $T$
 are summarized in Fig.~\ref{fig:phase-diagram}. 
For the parameter regimes shown, there are three distinct topological phases characterized by the
stroboscopic Chern numbers $C=0,\pm1$. 
TPTs occur at multiples of $T=\frac{\pi}{2}$\footnote{We remark that such vertical transition lines appear at multiples of $T=\frac{\pi}{2}$ beyond $T=\frac{3\pi}{2}$. For simplicity, and also because it is progressively harder to populate Floquet modes at larger periods in real settings, we focus on the regime $T < \frac{3\pi}{2}$.}.
These vertical TPTs can be obtained analytically from the Floquet operator, as shown in appendix~\ref{app:vertical-lines}.
From the phase diagram we can also discern a horizontal line at $J=1/3$, where the stroboscopic topological invariant becomes ill-defined.
Physically, this line corresponds to the case where all three hoppings in the driving protocol have the same strength.
As a consequence, the Floquet dynamics is momentarily frozen and time-reversal symmetry is restored.
As we will discuss later, because of symmetry enhancement, at $J=1/3$ the system exhibits nodal loop gap closures and non-propagating Majorana edge modes.

We remark that the stroboscopic invariant of Eq. ~\eqref{top-inv-strobo} can at times be insufficient to determine the correct number of edge modes at quasienergy $\epsilon=0,\pi$.
As shown in Ref.~\cite{Rudner:2013}, the correct formulation is given in terms of a pair of invariants $W_0$ and $W_{\pi}$, which incorporate the information about the \emph{micromotion} - \textit{i.e.} the full time evolution between periods, and about the probed band edge $\epsilon=0,\pi$:
\begin{align}
W_{\epsilon} &= \frac{1}{8 \pi^2} \int_0^T \mathrm{d}t \int_{-\pi}^{\pi} \mathrm{d} k_x  \int_{-\pi}^{\pi} \mathrm{d} k_y \nonumber \\
& \quad \times \Tr\left[ U_{\epsilon}^{-1} \partial_t U_{\epsilon} \left[ U_{\epsilon}^{-1} \partial_{k_x} U_{\epsilon}, U_{\epsilon}^{-1} \partial_{k_y} U_{\epsilon}\right]  \right]
\label{time-dep-top-inv}
\end{align}
Here, $U_{\epsilon}(\mathbf{k}; t)$ is an operator derived from $U(\mathbf{k}; t)$ by preserving the number of edge modes at $\epsilon$, which is smoothly connected to the identity at the end of the cycle, \textit{i.e.} $U_{\epsilon}(\mathbf{k}; T)=\mathds{1}$.
The stroboscopic Chern number is related to the time-integrated invariants as $C = W_{\pi} - W_0$~\cite{Rudner:2013, Yao:2017, Zhang:2020}.

A calculation of such time-integrated invariants at representative points in the three different phases yields the following.
In the $C=-1$ phase, we find $W_0=1$ and $W_{\pi}=0$. 
This indicates that this phase hosts only one edge mode at zero quasienergy.
In the $C=0$ phase, we obtain $W_0=1$ and $W_{\pi}=1$.
This indicates that this phase hosts \emph{one edge mode each} at zero and $\pi$ quasienergy, confirming that it is a Floquet anomalous phase - exhibiting edge modes despite a trivial bulk topological invariant.
In the $C=1$ phase, we find $W_0=0$ and $W_{\pi}=1$.
This phase is rather peculiar, because $W_0=0$ is not due to the absence of edge modes at zero quasienergy. 
Instead, as one can see upon inspecting the edge modes directly (see Appendix ~\ref{app:chiral-modes}), the $C=1$ phase hosts \emph{two} counter-propagating edge modes at zero quasienergy and one edge mode at $\pi$.


\subsection{Chirality tuning}
\label{subsec:chirality-tuning}


The three-step driving protocol inherently generates chiral Majorana modes propagating along the edges of a sample with finite size~\cite{Sato:2010, Po:2017, Fulga:2019}.
To verify this, we analyze semi-infinite strips with both armchair and zigzag edges and calculate the quasienergy spectrum and the corresponding eigenmodes as a function of the (one-dimensional) momentum $k$ defined for the infinite direction.
The superunit cell numbering used to construct the tight-binding Hamiltonians is presented in Fig.~\ref{fig:sketch}a),b).

\begin{figure}
\centering
\includegraphics[width=\columnwidth]{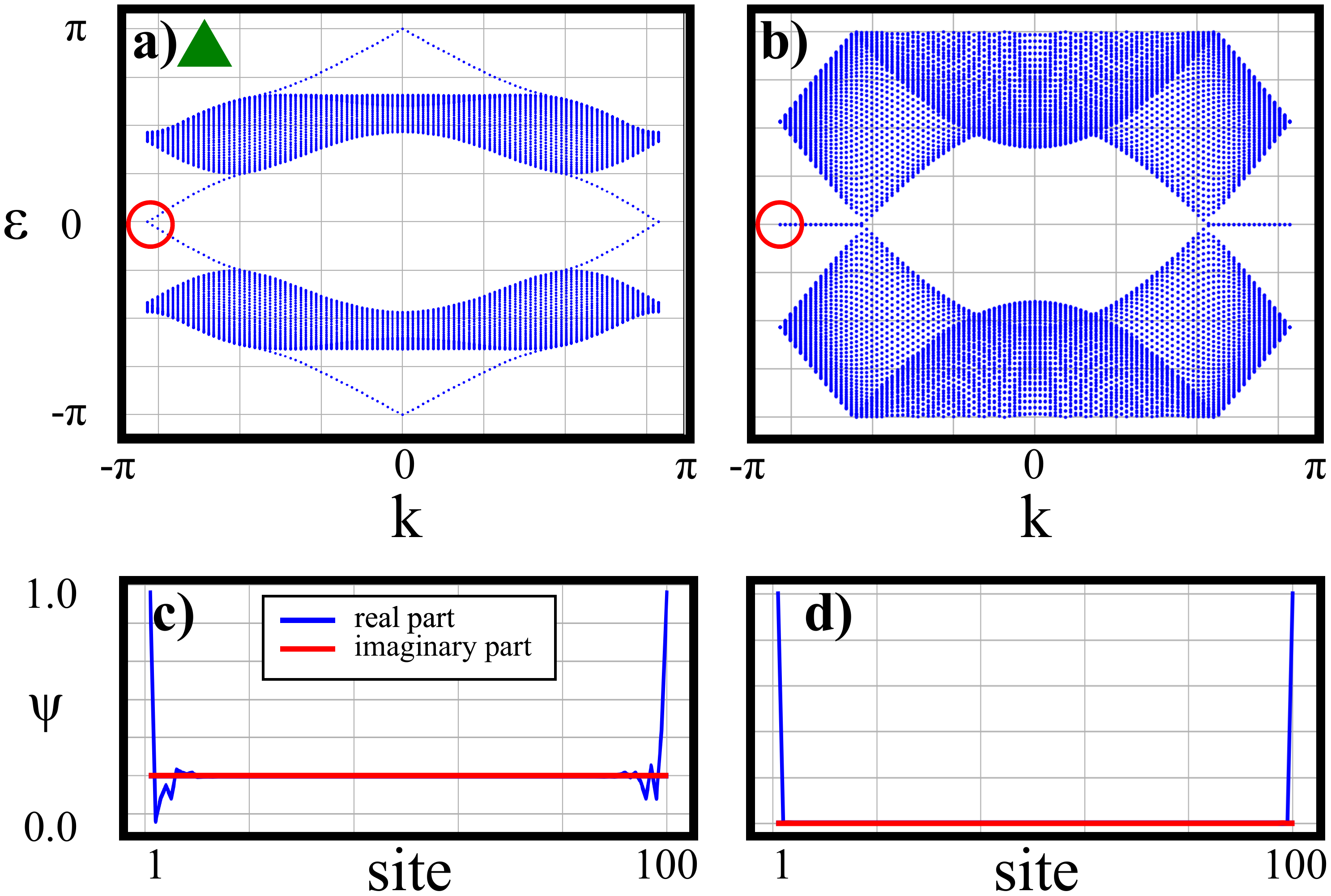}
\caption{Upper panels: quasienergy spectra as a function of momentum $k$ for semi-infinite strips with zigzag edges at $T=0.8$ and a) $J=0.8$ (chiral propagating MMs) and b) $J=1/3$ (non-propagating MMs).
By changing the value of $J$, it is possible to control the propagation velocity of the MMs.
Bottom panels: the corresponding eigenfunctions are purely real and localized at the edges.}
\label{fig:chirality-tuning}
\end{figure}

We find that the edge modes appear unsystematically for armchair geometries.
In zigzag geometries, they instead appear and disappear as a function of $T$ following the vertical pattern indicated in the phase diagram and matching the values of the topological invariants.
Furthermore, their wave functions at the one-dimensional momentum $k=0, \pm\pi$ is purely real, indicating Majorana character (see Fig.~\ref{fig:chirality-tuning}c), d)).
This is consistent with previous works dealing with similar driving protocols~\cite{Po:2017, Fulga:2019}.
In appendix \ref{app:chiral-modes}, we present a detailed analysis of the edge modes in each phase.

Because of the finite slope in their dispersion, as shown clearly in Fig.~\ref{fig:chirality-tuning}, the MMs in driven zigzag geometries represent \emph{chiral} edge states that propagate along the boundaries of the sample with finite and opposite velocities $\frac{\mathrm{d} \epsilon}{\mathrm{d} k}$,
 in contrast to  immobile MMs in the static system.
The chirality  emerges from an appropriate  choice of driving protocol that breaks time-reversal symmetry, allowing the MMs to propagate.
This mechanism is an alternative to the use of magnetic fields in quantum anomalous Hall insulators coupled to $s$-wave superconductors~\cite{Qi:2010, Chung:2011, Wang:2015, He:2017, Lian:2018, Hoegl:2020} or in $p\pm ip$ superconducting topological insulators~\cite{Chiu:2018, WangPRL:2018, He:2019}.

Note that  at $J=1/3$ the quasienergy dispersion  (see  Fig.~\ref{fig:chirality-tuning}b) exhibits completely flat, gapped bands at $\epsilon=0$.
and  exponentially edge localized  as evinced from Fig.~\ref{fig:chirality-tuning}d).
By traversing the $J=1/3$ line in parameter space,  the chirality of the MMs can thus be reversed.
In other words, the chirality is governed by the sequence in the driving protocol~\cite{Rudner:2013}.
For instance, while we have explicitly considered the sequence $1 \to 2 \to 3$ in the static phase diagram (Fig.~\ref{fig:sketch}c)), the opposite order $1 \to 3 \to 2$ yields the same quasienergy spectra, but with edge modes having opposite chirality.  
Note that at $J=1/3$, the system recovers an emergent time-reversal symmetry.
This renders the two drive sequences equivalent at $J=1/3$.
Consequently, it is possible to switch chiralities by reversing the drive sequence across $J=1/3$.

We note that the behavior at $J=1/3$ is an instance of \emph{frozen dynamics}, a phenomenon that is known to give rise to topological phase transitions at non high-symmetry points in periodically driven topological chains~\cite{Molignini:2018,Molignini:2020-multifrequency}.
While previously conjectured only for 1D systems, we establish here that frozen dynamics is more ubiquitous and can also be associated with more complex topological phase transitions.
Though $\mathcal{H}(t)$ at $J=1/3$  seems equivalent to its static counterpart at $J_x=J_y=J_z$,  the physics is determined by  the effective hamiltonian $\mathcal{H}_{\text{eff}}$ which is expected to be different, because for general matrices $\log \exp A \neq A$.
Furthermore, as we will explain in detail in the next section, at $J=1/3$ the bulk dispersion develops gap closures --- both at $0$ and $\pi$ --- in the form of nodal loops.

\section{Dirac and nodal loop gap closures in periodically driven Kitaev model}
\label{sec:NL}

\subsection{ Correlation function and fidelity susceptibility for periodically driven Kitaev model \label{sec:Majorana_Wannier_fidelity_driven}}

In this section, we demonstrate that the quantum criticality of the periodically driven Kitaev model is also well described by the Majorana-Wannier state correlation function and fidelity susceptibility discussed in Sec.~\ref{sec:static_Majorana_Wannier_fidelity}, although in a slightly different form. 
We start by considering the Wannier state in Eq.~(\ref{Wannier_static}) constructed from the lowest band stroboscopic Bloch eigenstate $|u_{-}({\bf k})\rangle$ of the effective Hamiltonian in Eq.~(\ref{Heff_strobo}).
One observes that the Fourier transform of the stroboscopic Berry curvature yields a correlation function that measures the overlap between the Wannier states~\cite{Wang06,Marzari12,Gradhand12,Chen:2017,Chen-Sigrist-book:2019,Chen-Schnyder:2019}, sandwiching the operator $R^{x}\hat{y}-R^{y}{\hat x}$,
\begin{eqnarray}
\tilde{F}_{2D}({\bf R})&=&\int\frac{d^{2}{\bf k}}{(2\pi)^{2}}F({\bf k})e^{i{\bf k\cdot R}}=-i\langle{\bf R}|R^{x}\hat{y}-R^{y}{\hat x}|{\bf 0}\rangle
\nonumber \\
&=&-i\int d^{2}{\bf r}(R^{x}\hat{y}-R^{y}{\hat x})W^{\ast}({\bf r-R})W({\bf r}),
\label{Majorana_Wannier_corr_driven}
\end{eqnarray}
where $W({\bf r-R})=\langle{\bf r}|{\bf R}\rangle$ is the Wannier function in the Majorana basis. Secondly, following the recipe in Sec.~\ref{sec:static_Majorana_Wannier_fidelity}, we construct the fidelity susceptibility by considering the behavior of the stroboscopic Bloch eigenstate $|u_{-}({\bf k})\rangle$ as moving 
\newpage

\onecolumngrid

\begin{figure}
\centering
\includegraphics[width=0.9\textwidth]{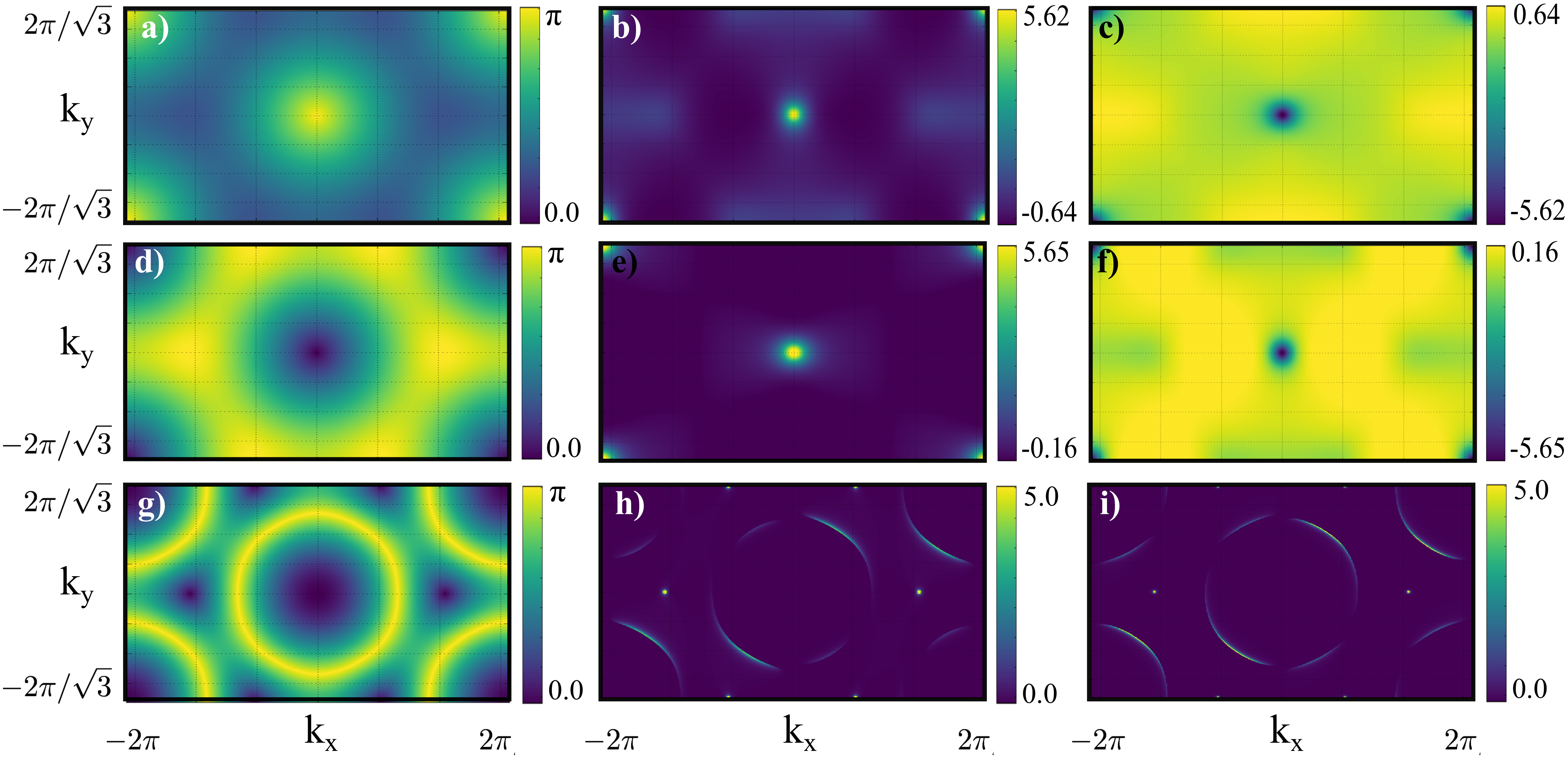}
\caption{Illustration of the quasienergy dispersion at the TPTs (left column) and the stroboscopic curvature function right before and after (middle and right columns) in the driven Kitaev model.
a)-c) Change in chiral $\pi$ modes across $T=0.5\pi$.
d)-f) Change in chiral zero modes across $T=1.0\pi$.
g)-i) Nodal loop gap closures and corresponding divergences in the curvature function across $J=1/3$ at $T=0.75\pi$.}
\label{fig:curv-func-Kitaev-honeycomb-driven}
\end{figure}

\twocolumngrid

\noindent
from momentum ${\bf k}$ to ${\bf k}+\delta{\bf k}$ along any direction. The rotation of $|u_{-}({\bf k})\rangle$ in the Hilbert space along this trajectory $|\langle u_{-}({\bf k})|u_{-}({\bf k}+\delta{\bf k})\rangle|=1-g_{\mu\nu}\delta k^{\mu}\delta k^{\nu}/2$
defines the quantum metric 
\begin{eqnarray}
g_{\mu\nu}&=&\frac{1}{2}\langle\partial_{\mu}u_{-}|\partial_{\nu}u_{-}\rangle
+\frac{1}{2}\langle\partial_{\nu}u_{-}|\partial_{\mu}u_{-}\rangle
\nonumber \\
&-&\langle\partial_{\mu}u_{-}|u_{-}\rangle\langle u_{-}|\partial_{\nu}u_{-}\rangle.
\end{eqnarray}

For the effective Hamiltonian in Eq.~(\ref{Heff_strobo}) of the form ${\cal H}_{\rm eff}({\bf k})={\bf d}({\bf k})\cdot{\boldsymbol\sigma}$, it can be verified that\cite{Panahiyan20}
\begin{eqnarray}
g_{\mu\nu}=\frac{1}{4}\partial_{\mu}\hat{\bf d}\cdot\partial_{\nu}\hat{\bf d}
=\frac{1}{4d^{2}}\left(\sum_{i=1}^{3}\partial_{\mu}d_{i}\partial_{\nu}d_{i}-\partial_{\mu}d \partial_{\nu} d \right),\;\;\;
\end{eqnarray}
with ${\hat{\bf e}}_{\mu}=\partial_{\mu}\hat{\bf d}/2$ playing the role of a vierbein and $d=\sqrt{d_1^2+d_2^2+d_3^2}$.
By treating $|u_{-}({\bf k})\rangle$ as a two-component spinor and momentum ${\bf k}$ as a tuning parameter, the determinant of the quantum metric serves as a representative local fidelity susceptibility\cite{You07,Zanardi07,Gu08,Gu10,Kolodrubetz17} (also called geometric orbital susceptibility\cite{Piechon16})
\begin{eqnarray}
g=\det\,g_{\mu\nu}=\frac{1}{4}F({\bf k})^{2}\equiv\chi_{F}, 
\label{fidelity_sus_driven}
\end{eqnarray}
which is found to be equal to the square of the stroboscopic Berry curvature $F({\bf k})$ for any parametrization of ${\bf d}({\bf k})$\cite{Ma13,Ma14,Panahiyan20}. 


We now detail the first type of TPT occurred in the periodically driven Kitaev model. The investigation starts from the behavior of the quasienergy dispersion and of the corresponding stroboscopic curvature function as shown in Fig.~\ref{fig:curv-func-Kitaev-honeycomb-driven}.
From panels a)-f), depicting the behavior across the vertical TPTs at $T=0.5\pi$ and $T=1.0\pi$, we recognize that the quasienergy gap closures are conical, leading to the stroboscopic Berry curvature $F({\bf k})$ in Eq.~(\ref{top-inv-strobo}) having a 2D Lorentzian shape
\begin{eqnarray}
F({\bf k}_{0}+\delta{\bf k})=\frac{F({\bf k}_{0})}{1+\xi_{x}^{2}\delta k_{x}^{2}+\xi_{y}^{2}\delta k_{y}^{2}}, 
\label{Ornstein_Zernike}
\end{eqnarray}
where $\delta{\bf k}$ is a small displacement along any direction. The critical behavior of the stroboscopic Berry curvature is such that the Lorentzian peak of Eq.~(\ref{Ornstein_Zernike}) gradually narrows and diverges, and flips sign as the tuning parameters ${\bf M}=(J,T)$ cross the critical point ${\bf M}_{c}$, as that described in Eqs.~(\ref{Fk0_xi_divergence}) and (\ref{Fk0_xi_exponent}). Numerically, we extract critical exponents
\begin{eqnarray}
\nu_{x}=\nu_{y}=1,\;\;\;\gamma=2,
\end{eqnarray}
which is consistent with the scaling laws $\gamma=\nu_{x}+\nu_{y}$ for 2D linear Dirac gap closures~\cite{Chen:2016, Chen-Sigrist:2016, Chen:2017, Kourtis17, Chen:2018, Molignini:2018, Chen-Sigrist-book:2019, Molignini:2019, Chen-Schnyder:2019, MoligniniReview:2019, Malard20_multicritical}.  Note that driving restores the full two dimensional nature of the TPT as opposed to TPTs in  the static model which belonged to the 1D universality class.  Furthermore, because of the Lorentzian shape in Eq.~(\ref{Ornstein_Zernike}), the Majorana-Wannier state correlation function $\tilde{F}_{2D}({\bf R})$ in Eq.~(\ref{Majorana_Wannier_corr_driven}) decays with the correlation length $\xi_{i}$ in the $i$ direction, justifying the designation of the critical exponent $\nu_{i}$ for $\xi_{i}$ in Eq.~(\ref{Fk0_xi_exponent}). Finally, the divergence of $F({\bf k}_{0})$ described by Eq.~(\ref{Fk0_xi_divergence}) implies that the fidelity susceptibility $\chi_{F}$ in Eq.~(\ref{fidelity_sus_driven}) also diverges at the HSP as the system approaches the critical point. Physically, this means the eigenstates $|u_{-}({\bf k}_{0})\rangle$ and $|u_{-}({\bf k}_{0}+\delta{\bf k})\rangle$ become more and more orthogonal as the system approaches the critical point. 
Because $F({\bf k}_{0})$ and $\chi_{F}$ basically share the same critical behavior, it justifies the usage of the exponent $\gamma$ for $F({\bf k}_{0})$.

\subsection{Nodal loop gap closure}

Besides the universality class of the 2D linear Dirac models at the vertical transition lines, Fig.~\ref{fig:phase-diagram} reveals  a transition line  at $J=1/3$.
This is the transition line where MMs lose their chirality in the finite-edge geometry, the Hamiltonian symmetries are enhanced, and the universality class changes. 
In $\mathbf{k}$-space, this line is characterized by the appearance of gap closures along 1D nodal loops.
This type of band inversion is radically different than the 2D Dirac-cone gap closures occurring at the vertical TPTs, when the number of MMs changes.
Similar 2D nodal-loop gap closures associated with emergent symmetries had been previously discovered only in periodically-driven Chern insulators~\cite{Molignini:2019}. 
Here, we demonstrate that these features are more ubiquitous than previously thought and appear to be a general feature of radical drives.

Panels g)-i) in Fig.~\ref{fig:curv-func-Kitaev-honeycomb-driven} illustrate the behavior of the system across the horizontal TPT at $J=1/3$.
In this case, both the quasienergy gap closures and the divergences in the curvature function follow closed loops in the Brillouin zone~\footnote{Because of frozen dynamics, the curvature function is also sensitive to the corresponding (conical) gap closures already known for the static model, as illustrated by the six bright dots in panels h) and i). In our work, we focus on the more exotic nodal loop gap closures.}.
The nodal loops are generated from the HSPs and progressively expand outward as a function of $T$ (cf. also Fig.~\ref{fig:comparison-nodal-loop-Kitaev-honeycomb-smaller}).
The nodal loop divergences are a clear sign of a different critical behavior belonging to a separate universality class.
It is also very interesting to note that this model accommodates quantum multicriticality: at $T=\frac{\pi}{2}n$, the nodal loops at $J=1/3$ coexist with the 0D Dirac gap closures already discussed in subsection \ref{sec:Majorana_Wannier_fidelity_driven}.

\subsection{Analytical derivation of nodal loops}

We now discuss the topological properties of the nodal loop gap closures in more detail.
At $J=1/3$, the Floquet stroboscopic operator becomes very easily tractable by directly exponentiating the Pauli matrices to yield
\begin{align}
U_{\mathbf{k}}(T,0) 
&= \cos \left( \frac{2T}{3} |\mathbf{v}| \right) \mathds{1}_2 - i \sin \left( \frac{2T}{3} |\mathbf{v}| \right) \hat{\mathbf{v}} \cdot \boldsymbol{\sigma},
\label{Floquet-operator-Kitaev-honeycomb-J-1-3}
\end{align}
where we have introduced the vector
\begin{align}
\mathbf{v} &=  \left( \begin{array}{c}
\sin\left( \mathbf{k} \cdot \mathbf{n}_1 \right) +  \sin\left( \mathbf{k} \cdot \mathbf{n}_2 \right) \\
\cos\left( \mathbf{k} \cdot \mathbf{n}_1 \right) +  \cos\left( \mathbf{k} \cdot \mathbf{n}_2 \right)  + 1 \\
0
\end{array}
\right),
\end{align}
whose magnitude can be computed by using some trigonometric identities to be
\begin{equation}
{\scriptstyle |\mathbf{v}| = 3 + 2 \left[  \cos ( \mathbf{k} \cdot (\mathbf{n}_1 - \mathbf{n}_2)) +   \cos (\mathbf{k} \cdot \mathbf{n}_1)  +  \cos ( \mathbf{k} \cdot \mathbf{n}_2)  \right]},
\end{equation}
and $\hat{\mathbf{v}}=\frac{\mathbf{v}}{|\mathbf{v}|}$. 
We can probe the gap closures by setting the diagonal (off-diagonal) elements of the Floquet operator to $\pm 1$ (zero), \textit{i.e.} $ \sin \left( \frac{2T}{3} |\mathbf{v}| \right) = 0$, $ \cos \left( \frac{2T}{3} |\mathbf{v}| \right) = \pm 1$.
This procedure leads to two equations that describe implicit curves in $k$-space where the quasienergy gap closes at $0$ or $\pi$:
\begin{align}
{\scriptstyle \sqrt{3 + 2 \left[  \cos ( \mathbf{k} \cdot (\mathbf{n}_1 - \mathbf{n}_2)) +   \cos (\mathbf{k} \cdot \mathbf{n}_1)  +  \cos ( \mathbf{k} \cdot \mathbf{n}_2)  \right]}} &= {\scriptstyle \frac{3\pi}{T} p_1}  \label{analytic-form-quasienergy-gap-0-nodal-Kitaev-honeycomb} \\
{\scriptstyle \sqrt{3 + 2 \left[  \cos ( \mathbf{k} \cdot (\mathbf{n}_1 - \mathbf{n}_2)) +   \cos (\mathbf{k} \cdot \mathbf{n}_1)  +  \cos ( \mathbf{k} \cdot \mathbf{n}_2)  \right]} } &= {\scriptstyle \frac{3\pi}{T} \left( \pi + 2\pi p_2 \right), } \label{analytic-form-quasienergy-gap-pi-nodal-Kitaev-honeycomb}
\end{align}
where $p_1$ and $p_2$ are integers.
The vectors $\mathbf{n}_1$, $\mathbf{n}_2$, and $\mathbf{n}_1 - \mathbf{n}_2$ span a triangle, such that the hexagonal structure and the inversion symmetry of the lattice will be preserved in the distribution of the gap closures in the quasienergy dispersion.

\begin{figure}
\centering
\includegraphics[width=0.9\columnwidth]{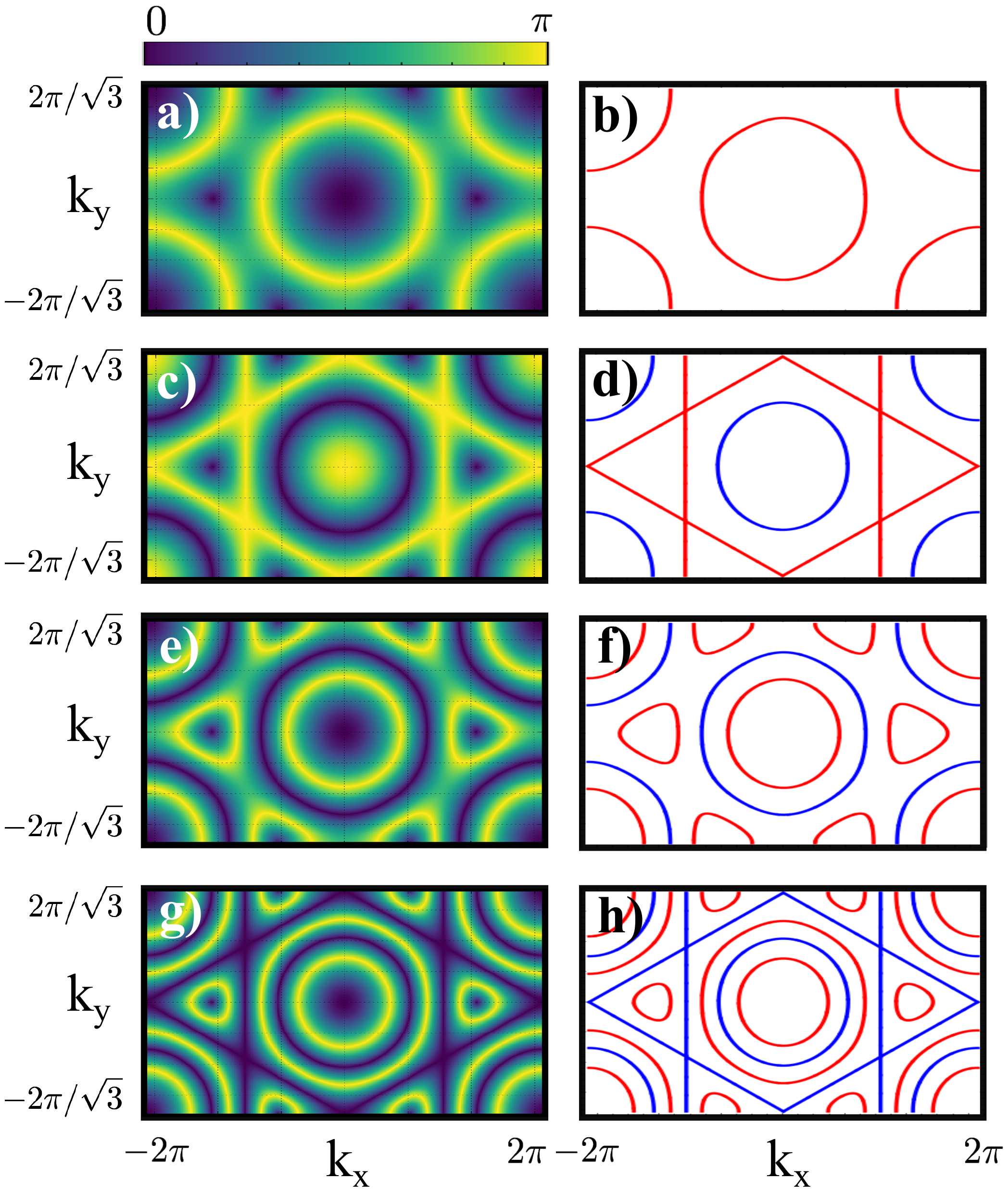}
\caption{Comparison of the quasienergy dispersion at $J=1/3$ (left panels) with the analytical prediction of the nodal loop gap closures (right panels) for various values of the driving period $T$: a)-b) $T=\pi$, c)-d) $T=\frac{3\pi}{2}$, e)-f) $T=\frac{5\pi}{2}$, g)-h) $T=3\pi$.}
\label{fig:comparison-nodal-loop-Kitaev-honeycomb-smaller}
\end{figure}

Fig.~\ref{fig:comparison-nodal-loop-Kitaev-honeycomb-smaller} shows a comparison between the quasienergy dispersion and the analytic form of the gap closures (Eqs. \eqref{analytic-form-quasienergy-gap-0-nodal-Kitaev-honeycomb} and \eqref{analytic-form-quasienergy-gap-pi-nodal-Kitaev-honeycomb} for $J=1/3$ and at various values of $T$.
We can clearly see that the analytical formulas precisely predict the location of the quasienergy gap closures.
These assume the form of highly complex patterns of nodal loops that are perfectly inversion and mirror symmetric with respect to the reciprocal lattice vectors, their sum, and their difference. 
Note that the nodal loops persist for all values $T>\frac{\pi}{2}$, not just integer multiples thereof.
Upon increasing the value of $T$ to very large values, the gap closures cascade to a plethora of nodal loops, as shown in Fig.~\ref{fig:comparison-nodal-loop-Kitaev-honeycomb-smaller}g) and h).

A natural question to ask is whether the gap closes in the same fashion across all nodal loops.
To address this question, we calculate the quasienergy dispersion analytically from Eq.~\eqref{eq:quasienergy-disp}, obtaining
\begin{align}
\theta(\mathbf{k}) &= \arccos \left( \cos \left( \frac{2T}{3} | \mathbf{v}(\mathbf{k})| \right) \right) = \frac{2T}{3} | \mathbf{v}(\mathbf{k})| \text{mod} 2\pi,
\end{align}
where $\text{mod} 2\pi$ means that the dispersion must be plotted within the first Floquet-Brillouin zone $[-\pi, \pi]$.

\begin{figure}
\centering
\includegraphics[width=0.9\columnwidth]{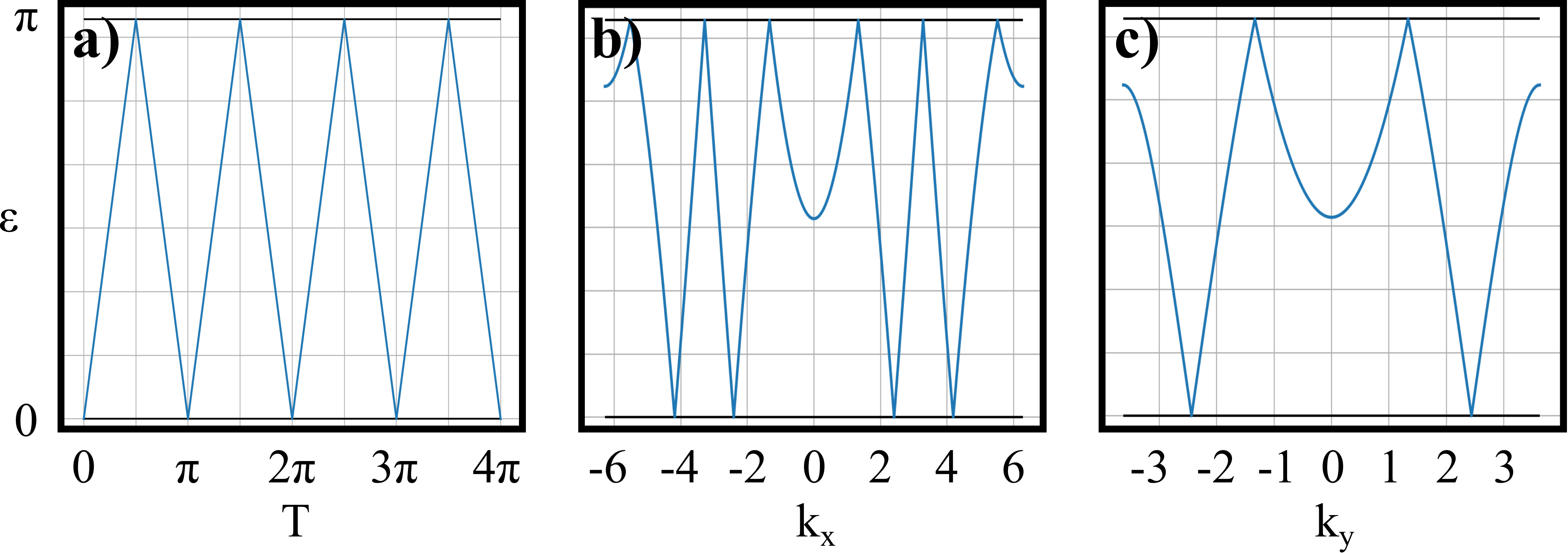}
\caption{Sections of the quasienergy dispersion across nodal loops, showing linear gap closures. a) $k_x=k_y=0$, b) $T=1.75\pi$, $k_y=0$, c) $T=1.75\pi$, $k_x=0$.}
\label{fig:gap-closures-NL-anal}
\end{figure}

With this analytical result, we can easily plot the behavior of the nodal loop gap closures as a function of $T$ and $\mathbf{k}$.
This is shown in Fig.~\ref{fig:gap-closures-NL-anal}.
Panel a) depicts the gap closures generated at $\mathbf{k}=(0,0)$ as $T$ is increased, while panels b) and c)  show the gap closures at fix $T$ for cuts along $k_x$ or $k_y$.
As we can see, the gaps always close and reopen linearly as a function of $T$, $k_x$ or $k_y$, both at quasienergy zero and $\pi$.
This is consistent across all the nodal loops generated in the system.
The linear gap-closure behavior can be understood as a consequence of the backfolding of the function $\frac{2T}{3} | \mathbf{v}|$ within the Floquet-Brillouin zone.

\subsection{Emergent time-reversal and mirror symmetries}

\begin{figure}
\centering
\includegraphics[width=\columnwidth]{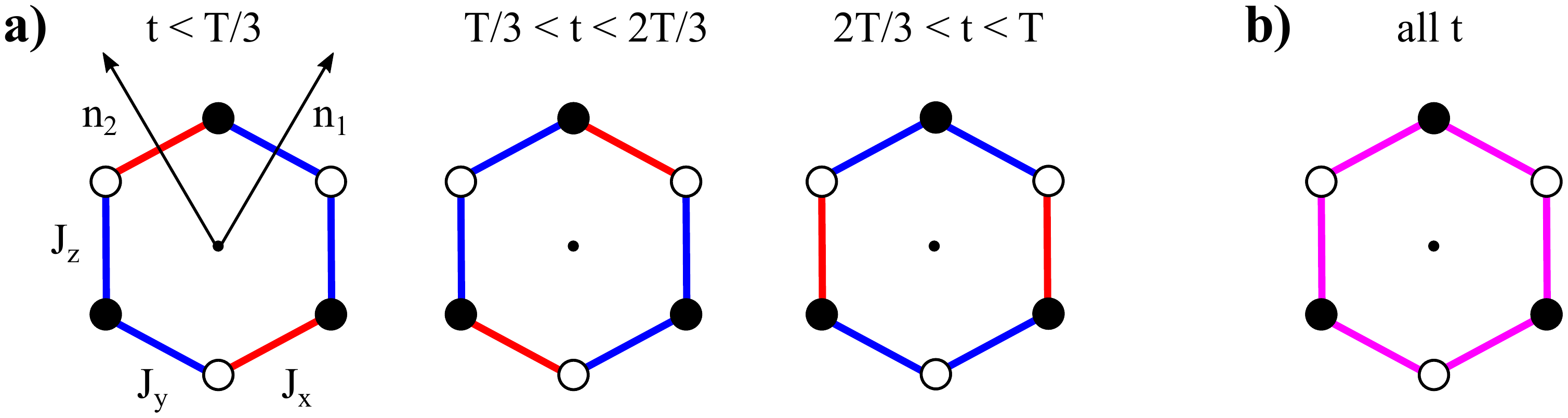}
\caption{a) Driving scheme at $J\neq 1/3$, where the normalization condition enforces two of the couplings to be equal.
At every driving step, the system is charge-conjugation (or equivalently inversion) symmetric.
b) Driving scheme at the nodal loop $J = 1/3$, where all the couplings are equal at every driving step. 
In this case, the effective dynamics is frozen and the driving realizes an effective Hamiltonian with additional mirror symmetry.}
\label{fig:symmetries-Kitaev-honeycomb}
\end{figure}

We now come back to the role symmetries in the driven system and how they are related to the different topological phases.
For any value of $J$, we have verified that the system obeys charge-conjugation symmetry with operator $\mathcal{C} = \sigma^0 \circ \mathcal{K}$ ($\mathcal{K}$ represents complex conjugation).
In the space of the stroboscopic Floquet Hamiltonian, this symmetry is equivalent to an inversion symmetry with operator $\mathcal{I}=\sigma^y$.
We can also appreciate this feature geometrically from Fig.~\ref{fig:symmetries-Kitaev-honeycomb}a), which shows that at each driving step the system is inversion symmetric with respect to the center of the hexagon.

We now examine how the symmetries are enhanced for the nodal loop transitions appearing at $J=1/3$.
As one can check directly by applying the transformations on \eqref{Floquet-operator-Kitaev-honeycomb-J-1-3} (or equivalently the corresponding $h_{\text{eff}}$), this TPT is associated with emerging time-reversal  and mirror symmetries, given by
\begin{align}
\mathcal{T} &= \sigma^z \circ \mathcal{K} \\
\mathcal{M}_x &= \sigma^0  \\
\mathcal{M}_y &= \sigma^y.
\end{align}
The geometric significance of the emerging mirror symmetry can be understood from Fig.~\ref{fig:symmetries-Kitaev-honeycomb}b).

Mirror symmetries are already known to stabilize nodal loop band inversions in static problems~\cite{Andreas-lecture-notes}.
In fact, their existence in combination with the non-spatial symmetries guarantees the topological protection of the nodal loop gap closures from certain mass terms in the effective stroboscopic description.
To understand this further, we note that the combination of time reversal and charge conjugation generates a chiral symmetry $\mathcal{S} = \sigma^z$, which prohibits any mass terms $M\sigma^z$.
We are then left with an effective hamiltonian of the form
\begin{equation}
h_{\text{eff}}^{J=1/3} = C[ \hat{v}_x \sigma^x + \hat{v}_y \sigma^y],
\end{equation}
where $C = -\frac{2T |\mathbf{v}|}{3}$.
This is in fact the form that we obtain from a direct calculation of $h_{\text{eff}}$ from the Floquet operator at $J=1/3$.
The mirror symmetry $\mathcal{M}_y$ acts as $\mathbf{k} \cdot \mathbf{n_1} \leftrightarrow -\mathbf{k} \cdot \mathbf{n_2}$ and therefore sends $v_x \to - v_x$ while it leaves $v_y$ unchanged. 
Therefore, any mass term $M \sigma^x$ is forbidden by the $\mathcal{M}_y$ symmetry, because $v_x + M \neq -v_x + M$.
The action of the mirror symmetry $\mathcal{M}_x$ gives instead no topological protection, because it acts as  $\mathbf{k} \cdot \mathbf{n_1} \leftrightarrow \mathbf{k} \cdot \mathbf{n_2}$  and leaves both $v_x$ and $v_y$ invariant.
Therefore, mass terms of the type $M \sigma^y$ are still allowed and the nodal loops emerging in the driven Kitaev model do not represent completely topologically protected band inversions.

\subsection{Universality class of nodal loop gap closures}

As a final analysis of the nodal-loop criticality, we determine their universality class by defining and extracting critical exponents from the stroboscopic curvature function of Eq.~\eqref{top-inv-strobo}.
We consider the stroboscopic curvature function as we approach the nodal loop transition vertically (as a function of $J$ for a constant $T$).
Because the nodal loop gap closures form ring-like structures around the origin, we need to first decide how to define diverging quantities and critical exponents.
For simplicity, we will focus only on circular loops, and ignore the hexagonal and ``diamond shaped'' loops that appear at larger $T$, although we have checked that the corresponding critical exponents are the same.

Approaching the nodal loop transition line, the stroboscopic curvature function displays ring-like divergences approaching the transition line, with diverging height and shrinking width.
For the circular nodal loops, we could then use polar coordinates $k = \sqrt{k_x^2 + k_y^2}$, $\phi = \arctan(k_y / k_x)$ to parametrize the contribution to the integral of the curvature function in a width $2\delta$ around this ring:
\begin{equation}
\mathcal{C}^{NL} = \int_{0}^{2\pi} \mathrm{d} \phi \int_{k_0 - \delta}^{k_0 + \delta} \mathrm{d}k \: F(k, \phi).
\end{equation}
Since the width of the Lorentzian ring is the same for all the values of $\phi$, we can approximate the above integral with a one-dimensional integral along a radial direction, which for simplicity we choose to be $k_x$ at fix $k_y=0$ (other directions are equivalent):
\begin{align}
\mathcal{C}^{NL} &\approx 2\pi k_0 \int_{k_x - \delta}^{k_x + \delta} \mathrm{d}k_x \: F(k_x, k_y=0) \nonumber \\
&\propto  \int_{k_x - \delta}^{k_x + \delta} \mathrm{d}k_x \: \frac{F_0}{1 + (k_x \xi)^2},
\end{align}
\textit{i.e.} we can reduce the problem to an effective one-dimensional one.
Then, conservation of $\mathcal{C}^{NL}$ as $J$ is brought closer to the critical value $J_c=1/3$ within the same phase should imply a scaling law
\begin{equation}
\nu = \gamma = 1
\label{eq:scaling-law}
\end{equation}
with
\begin{equation}
F_0(J) = \frac{1}{|J- J_c|^{\gamma}}, \qquad \xi(J) = \frac{1}{|J- J_c|^{\nu}}.
\end{equation}

\begin{figure}
\centering
\includegraphics[width=\columnwidth]{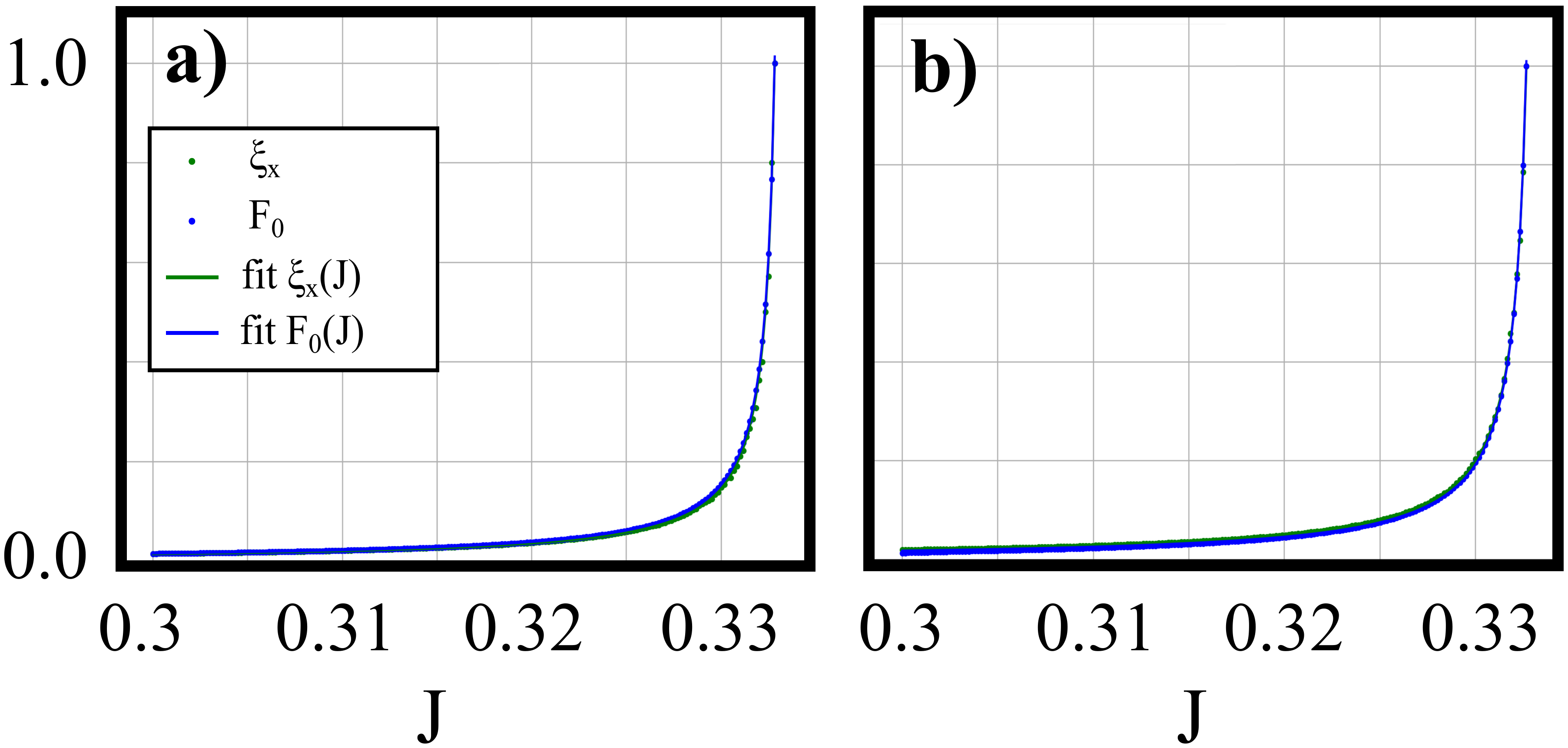}
\caption{Scaling behavior of the ring-divergences in the stroboscopic curvature function in correspondence with a nodal loop gap closure at a) quasienergy $0$ and b) quasienergy $\pi$.
The panels depict the behavior of the normalized height $F_0$ and inverse width $\xi$ of the curvature function for a cut at $k_y=0$.
All the quantities can be very well fitted by a curve $f(J) \propto \frac{1}{|J- J_c|}$.
Other types of cuts (not shown) lead to an equivalent precision of the fit.}
\label{fig:scaling-NL}
\end{figure}

To verify our analysis, we calculated the stroboscopic curvature function for two different nodal loops (one at quasienergy $0$, in Fig.~\ref{fig:scaling-NL}a), and the other at quasienergy $\pi$, in Fig.~\ref{fig:scaling-NL}b) and extracted its height and width (FWHM) as a function of $J$ for a cut at $k_y=0.0$ (other cuts lead to equivalent results).
The fits of these quantities indicate an excellent agreement with the scaling law \eqref{eq:scaling-law}, for both kinds of nodal loops.

We finally comment on the connections between 1D and 2D physics portrayed in this work.
Essentially, there are two mechanisms for dimensional reduction at play in the 2D Kitaev model: 1) the mapping from 2D to 1D at the static level, and 2) the effective 1D physics of the nodal loop gap closures in the Floquet Hamiltonian.
The mathematical connection from a 2D Kitaev model to an effective 1D chain is obtained by writing the Heisenberg equations of motion for the Majorana operators of the honeycomb model, and realizing that they describe a family of 1D Kitaev Hamiltonians where the chemical potential becomes k-dependent~\cite{Thakurathi:2014}.
This mapping does not change the symmetry class. 
The 1D Kitaev model hosts TPTs described by a Dirac low-energy theory~\cite{Molignini:2018}, which is exactly the same type of low-energy theory exhibited by the TPTs in the 2D Kitaev model. 
The 1D scaling laws (with only one radial scaling direction) exhibited by the nodal loop gap closures are instead of a different kind. 
These appear because the nodal loops, being extended gap closures with a uniform ``shrinking'' property, effectively behave as 1D objects, and have to be contrasted with the Dirac low-energy theories mentioned above, where the gap closes at 0D points in k-space and therefore we need two directions to define a scaling. In this case, symmetries are important. 
Nodal loops appear only when additional symmetries constrain the effective low-energy theory to not encompass the full Pauli space (for instance when it is defined in terms of two Pauli matrix instead of three like in this problem). 
These could be different chiral symmetries, or the combination of mirror symmetries with time-reversal like in the model considered in this work.

\section{Conclusions and Outlook}
\label{sec:outlook}

Based on the curvature function that integrates to the topological invariant, we have introduced Majorana-Wannier state correlation functions and momentum-dependent fidelity susceptibilities to characterize TPTs in both the static and periodically driven Kitaev model on the honeycomb lattice.
These quantities ubiquitously define critical exponents $\nu$ and $\gamma$ with respect to any static or dynamic tuning parameters, from which the universality class can be determined. 
Our analysis reveals the existence of cross-dimensional universality classes both in static and driven settings.
While being defined in a 2D geometry, the static model belongs in fact to the universality class of 1D Dirac models.
This result is consistent with previous findings highlighting that the 2D Kitaev model can be rewritten as a family of 1D Kitaev chains in the vortex-free sector~\cite{Thakurathi:2014}.
We note that we only focussed on the vortex-free sector which is preserved by the driving scheme, and extensions to the vortex-full sector are left for future investigations.

Periodic driving can be used to engineer multiple and coexisting 1D and 2D universality classes in the same system.
For the three-step driving protocol we consider, the stroboscopic and time-integrated topological invariants indicate a sequence of TPTs where \emph{chiral} MMs are created/annihilated. 
As a result, the system hosts both the universality class of prototype 2D Dirac models, and a 1D-like nodal loop type of transition owing to the emergent time-reversal and mirror symmetries when all the driving steps become equal.
Moreover, the restoration of time-reversal symmetry makes the MMs in the nodal loop phase nondispersive. 
By carefully tuning in and out of such parameter region, it is therefore possible to control the propagation velocity of the MMs, and even swap their chirality by switching to the reversed driving protocol.

Our results suggest that periodic driving offers an alternative to the use of magnetic fields to break time-reversal symmetry, and consequently manipulate the universality class and control the chirality of MMs. 
As a future question to explore, it would be interesting to introduce an imbalance between all the couplings $J_i$ at each step. 
This perturbation is expected to impact the TPTs, as it preserves charge-conjugation symmetry but destroys the mirror symmetries.
Another direction to pursue is the extension to generalized Kitaev models hosting parafermionic excitations~\cite{Barkeshli:2015}.
We also remark that the existence of nodal loop gap closures was recently confirmed in experimental realizations of similar periodically-driven honeycomb lattices in ultracold atoms~\cite{Wintersperger:2020}.
The possibility of realizing nodal loop gap closures in these settings could allow for a direct verification of their exotic topology and universality class, either by state tomography~\cite{Flaescher:2016} or by continuously or suddenly quenching across the TPT~\cite{Liou:2018}.

\acknowledgments

We kindly acknowledge financial support by the ETH Z\"{u}rich Foundation and Giulio Anderheggen.
This work is partially funded by EPSRC Grants No. EP/P009565/1 and by the European Research Council under the European Union's Seventh Framework Programme (FP7/2007-2013)/ERC Grant Agreement No. 319286 Q-MAC, and the productivity in research fellowship from CNPq. 
Albert Gasull acknowledges support from the International Max-Planck Research School for Quantum Science and Technology (IMPRS-QST).
Computation time on the ARCUS cluster of the University of Oxford and Euler cluster of ETH Zurich is gratefully acknowledged.
The authors would like to thank Monika Aidelsburger for fruitful discussions.

\appendix

\section{Analytic expression for the Floquet operator}
\label{app:anal-UF}

For the three-step drive presented in the main text, the resulting Floquet operator can be computed as
\begin{align}
U_F(\mathbf{k}) &= \exp \left( -i |\mathbf{v}_1| \frac{\hat{\mathbf{v}}_1}{|\mathbf{v}_1|} \cdot \boldsymbol{\sigma} \right)
\exp \left( -i |\mathbf{v}_2| \frac{\hat{\mathbf{v}}_2}{|\mathbf{v}_2|} \cdot \boldsymbol{\sigma} \right) \nonumber \\
&\times
\exp \left(-i |\mathbf{v}_3| \frac{\hat{\mathbf{v}}_3}{|\mathbf{v}_3|} \cdot \boldsymbol{\sigma} \right)
\end{align}
with the vectors
\begin{align}
\mathbf{v}_1 \equiv \frac{2T}{3} \left( \begin{array}{c} 
J \sin(\mathbf{k} \cdot \mathbf{n}_1) +  \frac{1-J}{2} \sin(\mathbf{k} \cdot \mathbf{n}_2) \\
J \cos(\mathbf{k} \cdot \mathbf{n}_1) +  \frac{1-J}{2} \cos(\mathbf{k} \cdot \mathbf{n}_2) + \frac{1-J}{2} \\
0 \end{array} \right), \\
\mathbf{v}_2 \equiv \frac{2T}{3} \left( \begin{array}{c} 
\frac{1-J}{2} \sin(\mathbf{k} \cdot \mathbf{n}_1) +  J \sin(\mathbf{k} \cdot \mathbf{n}_2) \\
\frac{1-J}{2} \cos(\mathbf{k} \cdot \mathbf{n}_1) +  J \cos(\mathbf{k} \cdot \mathbf{n}_2) + \frac{1-J}{2} \\
0 \end{array} \right), \\
\mathbf{v}_3 \equiv \frac{2T}{3} \left( \begin{array}{c} 
\frac{1-J}{2} \sin(\mathbf{k} \cdot \mathbf{n}_1) +  \frac{1-J}{2} \sin(\mathbf{k} \cdot \mathbf{n}_2) \\
\frac{1-J}{2} \cos(\mathbf{k} \cdot \mathbf{n}_1) +  \frac{1-J}{2} \cos(\mathbf{k} \cdot \mathbf{n}_2) + J \\
0 \end{array} \right).
\end{align}
By utilizing the formula for the exponential of Pauli matrices, $\exp \left( -i\lambda \hat{\mathbf{v}} \cdot \boldsymbol{\sigma} \right) = \cos \lambda \mathds{1} - 1 \sin \lambda \hat{\mathbf{v}} \cdot \boldsymbol{\sigma}$, we can multiply out the three terms.
After some tedious algebra, we arrive at the following result
\begin{equation}
U_F(\mathbf{k}) = \left( \begin{array}{cc}
A(\mathbf{k}) & B(\mathbf{k}) \\
-B^*(\mathbf{k}) & A^*(\mathbf{k})
\end{array}
\right),
\end{equation}
where the entries are given by
\begin{widetext}

\begin{align}
A(\mathbf{k}) &= 
\scriptstyle{
\cosh \left( \frac{T}{3} \sqrt{2 - 4J + 6J^2 + 2(-1+J)^2 \cos(k_x) - 8J(-1 + J) \cos \left( \frac{k_x}{2} \right) \cos \left( \frac{\sqrt{3}k_y}{2} \right) } \right)} \nonumber \\
& \scriptstyle{ \quad \times \cosh \left( \frac{T}{3} \sqrt{2 - 4J + 6J^2 +4(1-J) \cos(k_x) +2(1 - J^2) \cos \left( \frac{k_x}{2} \right) \cos \left( \frac{\sqrt{3}k_y}{2} \right) +2(1-J)(1-3J) \sin \left( \frac{k_x}{2} \right) \sin \left( \frac{\sqrt{3}k_y}{2} \right) } \right)} \nonumber \\
& \scriptstyle{\quad \times \cosh \left( \frac{T}{3} \sqrt{ \left(
-2J \cos \left( \frac{k_x - \sqrt{3} k_y}{2} \right) + (-1+J) (1 + \cos \left( \frac{k_x +\sqrt{3} k_y}{2} \right) \right)^2
+ \left(
2J \sin \left( \frac{k_x - \sqrt{3} k_y}{2} \right) + (-1+J) (1 + \sin \left( \frac{k_x +\sqrt{3} k_y}{2} \right) \right)^2
}
 \right)}
\end{align}
and
\begin{align}
B(\mathbf{k}) &= \scriptstyle{\frac{2T^3}{27} \left(1 - J + \exp \left( -\frac{i}{2} (k_x - \sqrt{3} k_y) \right) \left( \exp(i k_x ) (1 -J) + 2J \right) \right)
\left( -1 + J + \exp \left( -\frac{i}{2} \left( k_x - \sqrt{3} k_y \right) \right) \left( -1 + J - 2J \exp(i k_x) \right) \right)} \nonumber \\
& \scriptstyle{\quad \times
\left( -J \exp \left( \frac{i}{2}\sqrt{3} k_y \right) \left(-1 + J \right) \cos \left( \frac{k_x}{2} \right) \right) A(\mathbf{k}).} 
\end{align}

\end{widetext}

\section{Chiral edge modes}
\label{app:chiral-modes}

Here we analyze more in detail the stroboscopic edge modes existing in each phase of the phase diagram for semi-infinite strips with different edge geometries (zigzag and armchair -- see Fig.~\ref{fig:sketch}).
To obtain the quasienergy spectrum and the corresponding eigenmodes, we construct tight-binding Hamiltonians by following the superunit cell numbering presented in Fig.~\ref{fig:sketch}, and diagonalize the corresponding effective Floquet Hamiltonian \eqref{eff-Floquet-Ham} as a function of the (one-dimensional) momentum $k$ defined for the infinite direction.

\begin{figure}[h]
\centering
\includegraphics[width=\columnwidth]{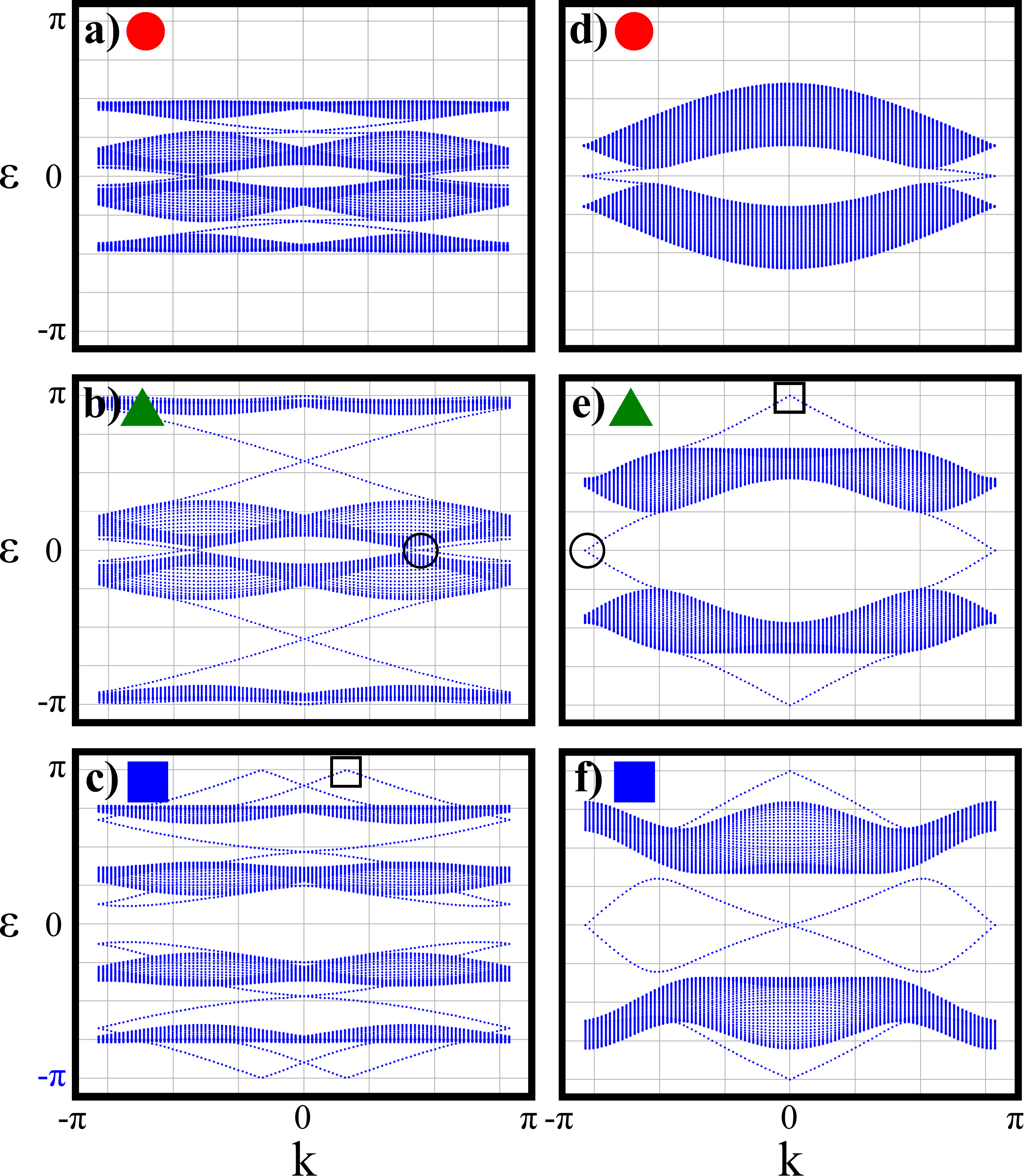}
\caption{Quasienergy dispersions as a function of momentum $k$ for semi-infinite strips with armchair edges (left panels) and zigzag edges (right panels). 
The colored symbols indicate the driving parameters in the phase diagram of Fig.~\ref{fig:sketch}.
The hollow circles and squares indicate the edge states plotted in Fig.~\ref{fig:chiral-eigenfunctions}.
All strips have $N=100$ sites in the finite direction.
}
\label{fig:chiral-quasienergies}
\end{figure}

The quasienergy dispersion as a function of the momentum $k$ for a few representative points in the $(J,T)$ phase diagram is shown in Fig.~\ref{fig:chiral-quasienergies}.
The bulk bands are separated by clear band gaps, but connected by pairs of in-gap states of opposite slopes.
While for zigzag geometries the edge states of opposite chirality always intersect at degenerate quasienergies $\epsilon=0$ or $\pm\pi$, armchair geometries show the existence of crossings at other quasienergies.
Because such edge states away from $\epsilon=0$ or $\pm\pi$ are energetically separated from their charge-conjugated partners at $-\epsilon$, they cannot hybridize to form a purely real wave functions and hence they cannot correspond to conventional Majorana edge states.
For this reason, we will instead focus on the states that cross $\epsilon=0$ or $\pm\pi$.
We have verified that all of such states are localized at the edges of the sample.

\begin{figure}
\centering
\includegraphics[width=\columnwidth]{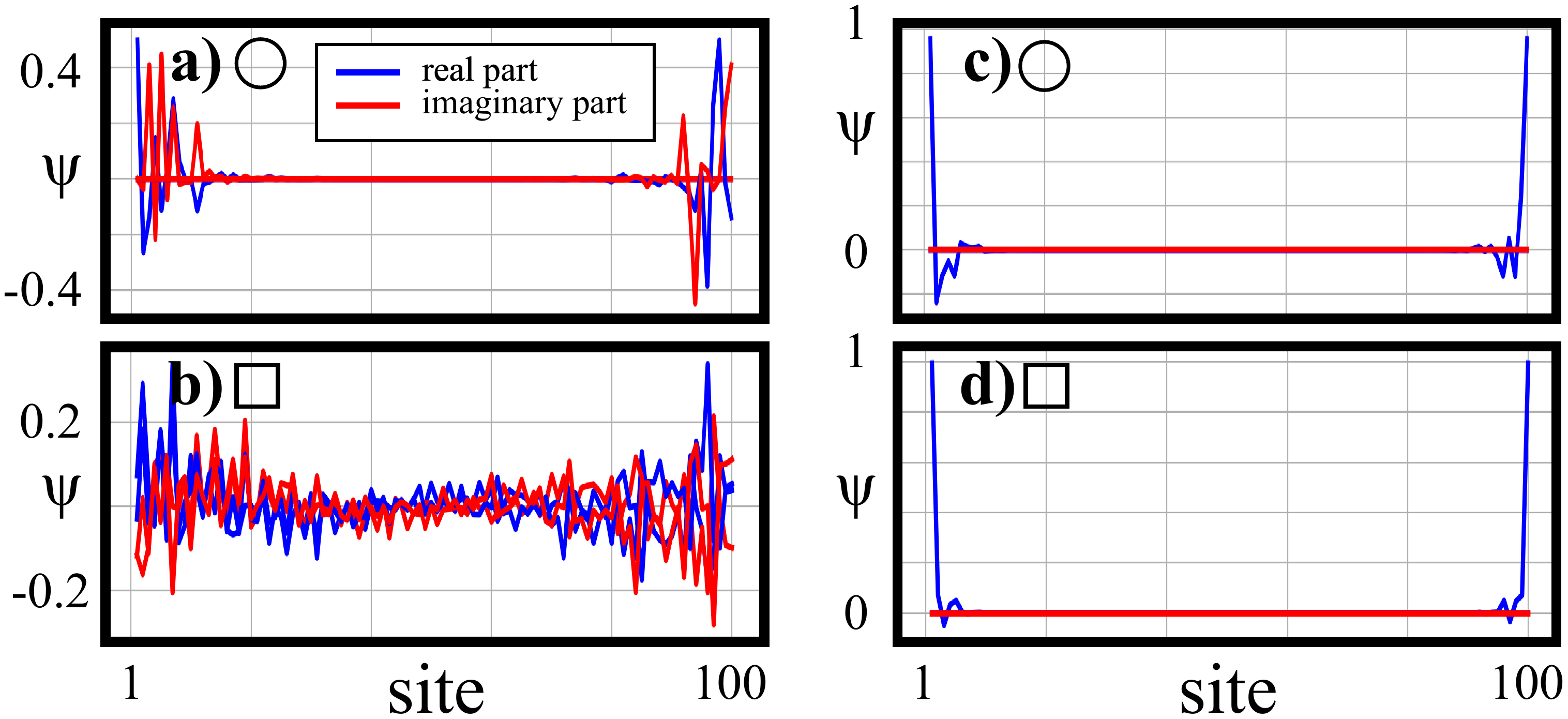}
\caption{Eigenfunctions for some edge states appearing in Fig.~\ref{fig:chiral-quasienergies}. Left (right) panels depict eigenfunctions for strips with armchair (zigzag) edges. The hollow circles and squares indicate which quasienergies they correspond to in Fig.~\ref{fig:chiral-quasienergies}.}
\label{fig:chiral-eigenfunctions}
\end{figure}

For zigzag geometries, we find that the creation and annihilation of edge modes perfectly reproduces the patterns found in the stroboscopic phase diagram.
Furthermore, the crossings are pinned at $k=0$ and $k=\pm$, where their wave function is purely real (see Fig.~\ref{fig:chiral-eigenfunctions} c) and d)), consistent with the definition of Majorana edge modes.
In particular, panel e) once again confirms the coexistence of zero and $\pi$ edge modes in the anomalous phases with $C=0$.
In panel f), we can instead recognize a pair of counter-propagating zero MMs.
Because these modes appear as floating bands within the gap, their topological invariant $W_0$ cancels out.
Analogous edge modes representing a weak topological phase and protected by particle-hole and translation symmetry have been observed for similar driving schemes~\cite{Fulga:2019}.
Static counterparts of such floating band modes were shown to lead to second-order topological superconducting states in the presence of $s_{\pm}$-wave superconductivity~\cite{Yan:2020}.

For armchair geometries, on the other hand, the edge modes appear unsystematically.
Furthermore, even when the in-gap modes intersect at quasienergy zero or $\pi$, their wave function is never purely real~\footnote{The intersections do not typically occur at $k=0$ or $k=\pi$, but we have verified that even when they do, the wave function is never purely real.}, and sometimes not even edge-localized (see Fig.~\ref{fig:chiral-eigenfunctions} a) and b)).
These states are thus indistinguishable from other complex edge states occurring at other arbitrary quasienergies and hence not necessarily topological. 
This is consistent with previous findings on similar systems -- for instance graphene -- where only the zigzag edges host topological edge modes.

\section{Analytical derivation of the vertical transition lines in the periodically driven Kitaev model}
\label{app:vertical-lines}

We can understand the periodic pattern of vertical TPTs by examining the form of the effective Hamiltonian, which by virtue of the piecewise constant drive can be calculated analytically.
We consider the case $J=1$ for simplicity, even though our derivation can be generalized to arbitrary values of $J$.
At $J=1$, the entries of the Floquet operator are computed to be
\begin{align}
{\scriptstyle A(\mathbf{k})} &= {\scriptstyle \cos^3 \left( \frac{2T}{3} \right)  - \left( e^{-i k_x} + e^{-i \mathbf{k} \cdot \mathbf{n_1}} + e^{-i \mathbf{k} \cdot \mathbf{n_2}} \right) \sin^2 \left( \frac{2T}{3} \right)  \cos \left( \frac{2T}{3} \right) } \label{entry-A-Floquet-op-Kitaev-honeycomb} \\
{\scriptstyle B(\mathbf{k})} &= {\scriptstyle - \left( 1 + e^{i \mathbf{k} \cdot \mathbf{n}_1} + e^{i \mathbf{k} \cdot \mathbf{n}_2} \right) \cos^2 \left( \frac{2T}{3} \right) \sin \left( \frac{2T}{3} \right) + e^{i k_x}  \sin^3\left( \frac{2T}{3} \right)}.
\end{align}
From this form we can extract the points in the phase diagram where we expect a quasienergy gap closure at $\theta=0$ or $\theta=\pi$, by demanding $B = 0$ and $A = \pm 1$.
From the first condition we obtain
\begin{equation}
\tan^2 \left( \frac{2T}{3} \right) = e^{-i \mathbf{k} \cdot \mathbf{e}_x} + e^{-i \mathbf{k} \cdot \mathbf{n}_2}  + e^{-i \mathbf{k} \cdot (\mathbf{n}_2 - \mathbf{e}_x)},
\label{condition-gap-closures-kitaev-honeycomb}
\end{equation}
which has a solution for the real variable $T$ only if the right-hand side is real.
This fixes a condition for the location of the gap closures in $k$-space, namely:
\begin{equation}
{\scriptstyle
\sin(k_x) = \sin \left( - \frac{k_x}{2} + \frac{\sqrt{3}}{2} k_y \right) = \sin \left( - \frac{3k_x}{2} + \frac{\sqrt{3}}{2} k_y \right) = 0,}
\end{equation}
which can be compactly summarized as
\begin{equation}
\mathbf{k}_0 = \left( \begin{array}{c}
m \pi \\
(2\tilde{m} -m) \frac{\pi}{\sqrt{3}} 
\end{array}
\right), \qquad m, \tilde{m} \in \mathbb{Z}.
\label{gap-closure-points-Kitaev-honeycomb}
\end{equation}
Note that these points encompass a larger set than the HSPs where the gap closes in the static system.
For these values, Eq. \eqref{condition-gap-closures-kitaev-honeycomb} becomes
\begin{equation}
\tan^2 \left( \frac{2T}{3} \right) = (-1)^m + (-1)^{\tilde{m}} + (-1)^{\tilde{m}-m}.
\end{equation}
Depending on the chosen integers $m$ and $\tilde{m}$, the right-hand side can be either $3$ (both $m$ and $\tilde{m}$ even) or $-1$ (all other possibilities). 
To avoid complex numbers when taking the square root to solve for $T$, only the positive number is a sensible solution.
Correspondingly, we obtain the analytic form for all the vertical TPTs:
\begin{equation}
\tan^2 \left( \frac{2T}{3}\right) = 3 \quad \Rightarrow \quad T = \frac{\pi}{2} n, \quad n \in \mathbb{N},
\label{analytical-TPTs-Kitaev-honeycomb}
\end{equation}
where we have accounted for the periodicity of the $\tan$ function to obtain all the other solutions.

We refine our analysis further by analyzing the structure of the entry $A$.
By inserting Eqs.~\eqref{gap-closure-points-Kitaev-honeycomb} and \eqref{analytical-TPTs-Kitaev-honeycomb} into Eq.~\eqref{entry-A-Floquet-op-Kitaev-honeycomb}, we can verify that it can indeed become $\pm 1$, depending on the integers $m$, $\tilde{m}$, and $n$:
\begin{align}
A(\mathbf{k}_0) =& {\scriptstyle \cos \left( \frac{2T}{3} \right) \left[  \cos^2 \left( \frac{2T}{3} \right) - \sin^2 \left( \frac{2T}{3} \right) \left( (-1)^m + (-1)^{\tilde{m}} + (-1)^{\tilde{m} - m}  \right) \right]} \nonumber \\
&= \begin{cases}
-1, \quad &\tilde{m}, m \in 2\mathbb{Z}, n \in 2\mathbb{Z}+1 \\
+1, \quad &\tilde{m}, m \in 2\mathbb{Z}, n \in 2\mathbb{Z} \\
\pm \frac{1}{2}, \quad &\tilde{m}, m \notin 2\mathbb{Z}.
\end{cases}
\end{align}
This formula demonstrates that at multiples of $T=\frac{\pi}{2}$, we will encounter a gap closure at the points $\mathbf{k} = (0,0)$ and $\mathbf{k} = (\pm 2\pi , \pm \frac{2\pi}{\sqrt{3}})$.
This analysis hence completely describes the characteristics of the stroboscopic topology already observed empirically from the behavior of the curvature function and the quasienergy dispersion.

\newpage
\bibliography{many-body-biblio}

\end{document}